\documentclass[aip,twocolumn,letterpaper]{revtex4}

\usepackage{epsfig}
\usepackage{xcolor}

\usepackage{graphics}
\usepackage{epsfig}

\usepackage[margin=1.0in]{geometry}

\begin{document}
\title{Constraints on the magnetic field evolution in tokamak power plants }
\author{Allen H Boozer}
\affiliation{Columbia University, New York, NY  10027 \linebreak ahb17@columbia.edu}

\begin{abstract}

Forty-five years ago a coordinate system was shown to exist that gave simple but exact expressions whenever and wherever a toroidal plasma equilibrium $\vec{\nabla}p=\vec{j}\times\vec{B}$ exists.  These coordinates, now called Boozer coordinates, which revolutionized the stellarator program, are also applicable to tokamaks.  Here expressions for Faraday's Law, the safety factor, and the internal inductance are derived.  Their constraints should be useful in the design of tokamak power plants and for the thoughtful allocation of resources to minimize the time and the cost to the achievement of practical fusion power.   Simple explanations are obtained for (1) why disruptions in tokamaks are so common, (2) why current-profile control though difficult may be required, especially during plasma shutdowns, and  (3) why only pulsed tokamaks seem possible.  Lack of familiarity with Boozer coordinates can make simple but exact expressions appear naive.  Complicated derivations with dubious assumptions have been interpreted as ``more rigorous.'' 

\color{black}

 \end{abstract}

\date{\today} 
\maketitle

%%%%%%%%%%%%%%%%%%%%%%%%%%%%%%%%%%%%%%%%%%%%%%%%%%%%%%%%%%

\section{Introduction}

Theory and computation  have three roles in the fusion program:  (1) Developing and employing of codes to make increasingly complete and reliable determination of both physics and engineering properties.  (2) Innovating ways to circumvent challenges to the development of fusion.  (3) Providing program leadership by clarifying what issues need to be addressed and how they can be addressed with minimal cost and time.   Unfortunately, the second and third of these roles have received inadequate attention with  a low funding priority in the public and private fusion programs.  

Here the third role will be discussed in the context of the tokamak program---in particular issues that separate tokamaks from stellarators.   The dominance of the external coils on the magnetic field structure in stellarators is fundamentally different from the plasma self-organization of tokamaks, in large part due to microturbulence, especially when fusion rather than external heating dominates.  The self-determined profiles of plasma properties include the profile of the plasma current, which determines the poloidal field in a tokamak.  

Optimization is an important part of the design of both tokamak and stellarator power plants.  Stellarator optimization is focused on the order of magnitude greater number of feasible external magnetic field distributions \cite{Boozer:RMP} than tokamaks.  Tokamak optimization explores the benefits of different plasma profiles.  The optimal choice for the external field can be accurately enforced, but the enforcement of the choice of optimal profiles is problematic.

 %%%%%%%%%%%%%%%%%%%%%%%%%

%%%%%%%%%%%%%%%%%%%%%% 
\begin{figure}
\centerline{ \includegraphics[width=3.2 in]{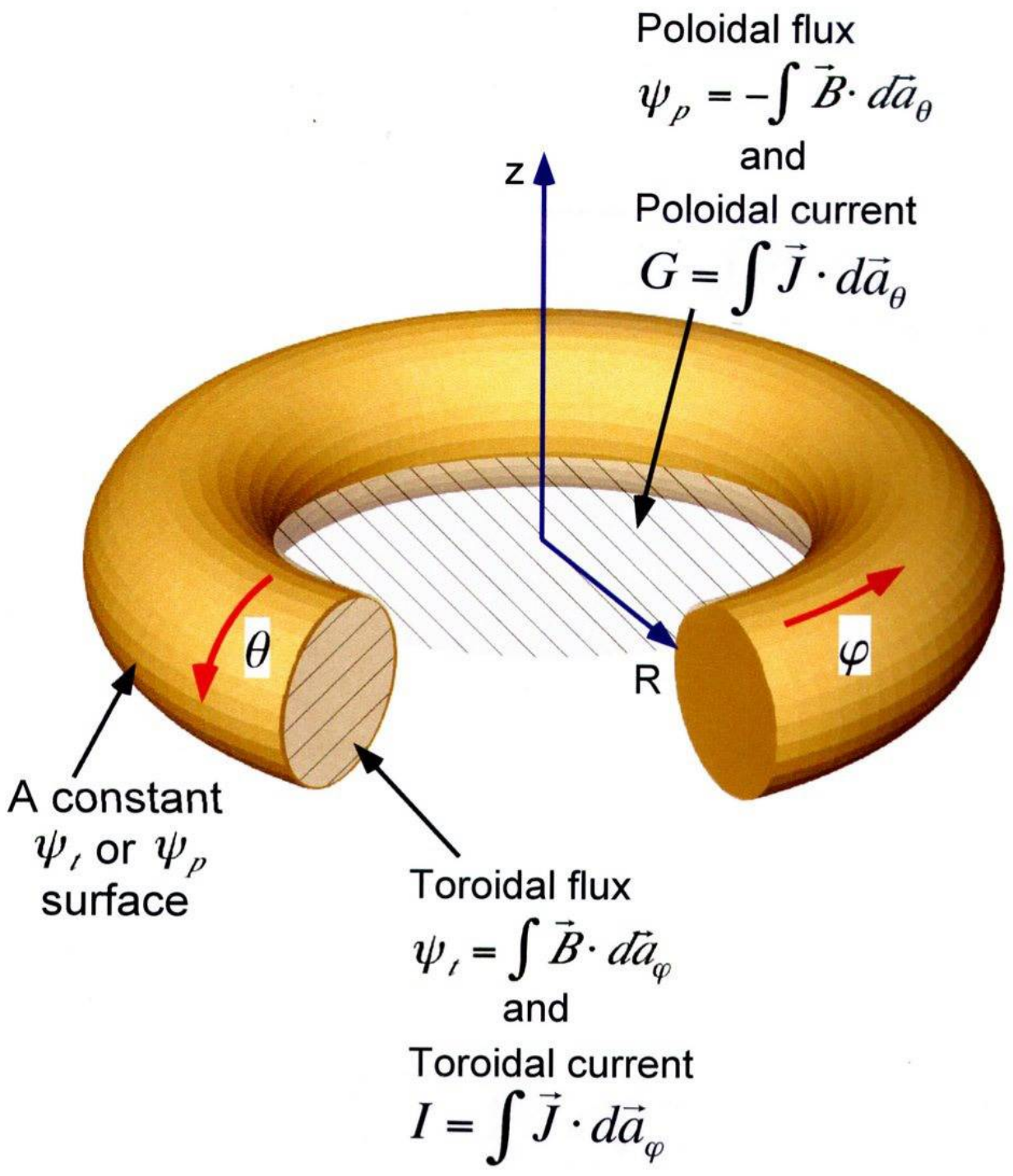}}
\caption{The toroidal flux $\psi_t$ is the magnetic flux enclosed by a toroidal magnetic surface. The poloidal flux $\psi_p(\psi_t,t)$ is  the magnetic flux going down through the hole in the center of the toroidal magnetic surface. The toroidal current is $I(\psi_t,t)$ is the current enclosed by the magnetic surface.  The poloidal current $G(\psi_t,t)$ is the current coming up through the hole in the center of the toroidal magnetic surface.  The current density in the figure is denoted by $\vec{J}$, but in this paper by $\vec{j}$.  This was Figure 1 in Boozer,  Nucl. Fusion \textbf{55}, 025001 (2015).  } 
\label{fig:fluxes-currents}
\end{figure}
%%%%%%%%%%%% 

Particularly problematic is the enforcement of a plasma current profile $I(\psi_t,t)/I_p(t)$.  $I(\psi_t,t)$ is the toroidal plasma current enclosed by a magnetic surface that encloses a toroidal magnetic flux $\psi_t$, Figure \ref{fig:fluxes-currents}, and $I_p(t)$ is the total plasma current.  

The plasma current profile is considered the primary determinant of whether a tokamak disrupts.  Disruptions are a sudden loss of plasma confinement and are caused by plasma instabilities: tearing modes, which destroy magnetic surfaces, kink modes, which helically deform the plasma, and loss of axisymmetric position control. They can also be caused by a radiative collapse, as when a piece of wall material falls into the plasma.  Disruptions will be discussed in Section \ref{Sec:disruption} and have been observed throughout the history of tokamaks and can affect the feasibility of fusion power plants \cite{Eiditis:2021}.  

\begin{figure}
\centerline{ \includegraphics[width=2.0 in]{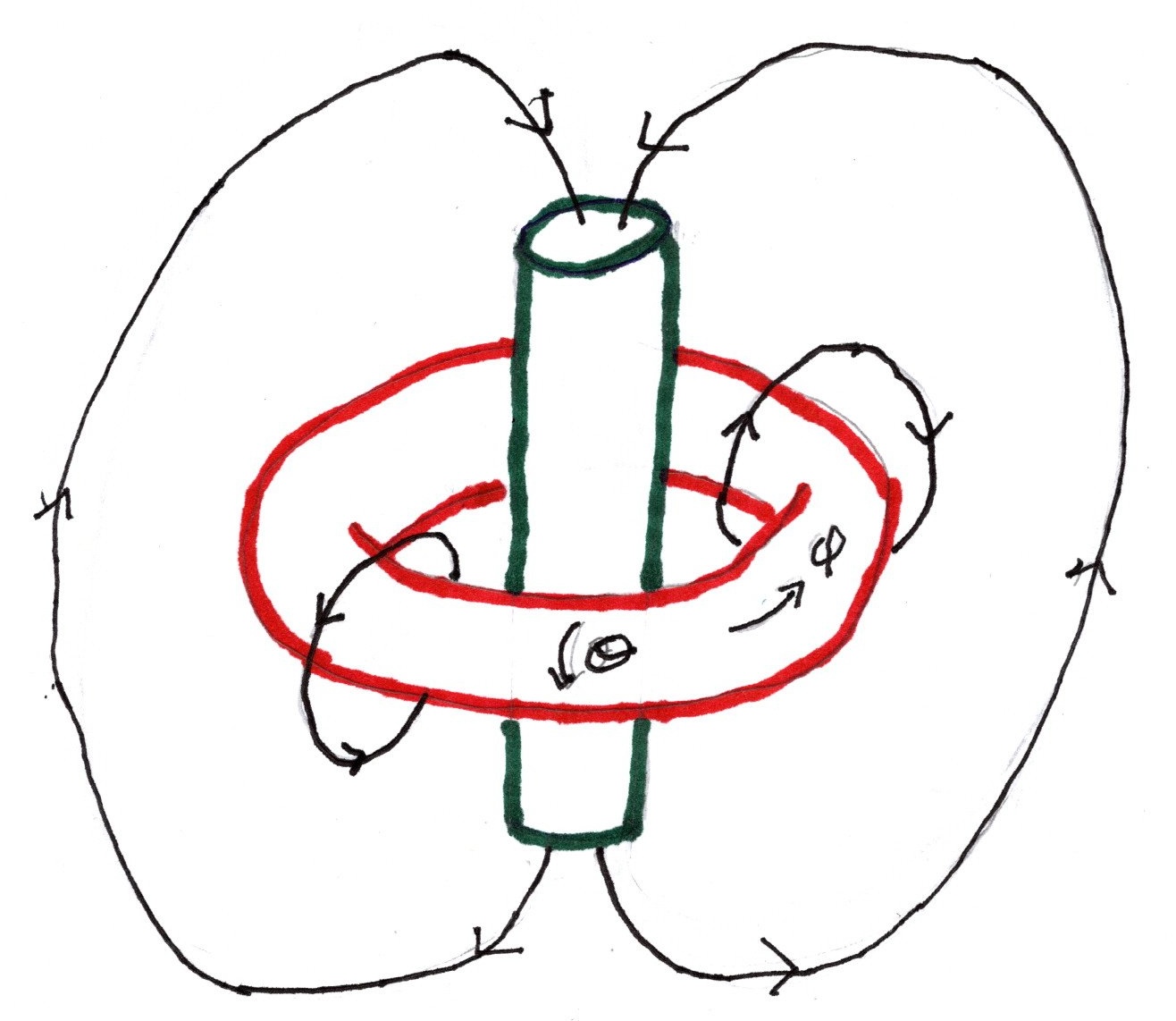}}
\caption{The lines of the poloidal magnetic field produced by the toroidal plasma current are shown together with the magnetic field produced by the central solenoid of a tokamak. }
\label{fig: B}
\end{figure}

As will be discussed in Section \ref{Sec:plasma maintenance},  the economic viability of power plants requires the total fusion power production be far larger than the power required to maintain the plasma.  For tokamaks, the only viable solution appears to be the induction of the plasma current by a central solenoid, Figure \ref{fig: B}.  The swing in the poloidal flux $\Psi_p^{sol}(t)$ that can be produced by the central solenoid is limited by basic engineering considerations to being only comparable to the total poloidal flux $\Psi_p^{pl}(t)$ produced by the plasma current.  This implies that not only do tokamaks appear to need to be pulsed but each pulse is limited to tens of minutes by the resistive decay of the induced flux $\Psi_p^{pl}$.

%%%%%%%%%%%%%%%%%%%%%%
%%%%%%%%%%%%

Deviations in profiles presumably also explain the degradation of the critical parameter for a fusion burn, $n\tau_ET$, with pulse length on a 10~second timescale, which is observed in tokamaks but not in stellarators \cite{Long pulse:2024}. \color{black}   Even more striking results from the W7-X stellarator are given in a 2026 paper \cite{Long puse:2026}. \color{black}  Despite the importance of the $n\tau_ET$ issue, here the focus is on  issues associated with Faraday's Law  and how they could be addressed in power-plant  design.  The reason is simple.  Important and unappreciated constraints are easier to derive from Faraday's Law.

Issues that are more difficult to solve and may be unresolvable for tokamak power plants are addressed by stellarators.  There are no fundamental power-plant issues for stellarators that are not also issues for tokamaks.  An envisioned advantage of tokamaks, especially spherical tokamaks, is a smaller unit size.   

The issues that separate tokamaks and stellarators are far more important for the feasibility of power plants than for the demonstration of deuterium-tritium (DT) ignition.  A ten second pulse followed by a disruption that does not produce extreme machine damage could be used to demonstrate DT ignition.  But, the feasibility of fusion power is highly dependent on ensuring disruptions are neither so violent nor so common that they require frequent replacement of neutron embrittled internal tokamak components.   It is difficult to argue with the statement at the beginning of the abstract of a paper \cite{Eiditis:2021} by Nicholas Eidietis: ``Disruptions present a great challenge to achieving an economically viable commercial tokamak fusion reactor. Disruption handling, including prevention, mitigation, and resilient design, must be incorporated into future reactor designs at the same priority as core performance and steady-state heat flux removal."

The defining issue for stellarators was the absence of a continuous symmetry, which long precluded the design of fusion-relevant stellarators.  Arguments for the unsuitability of stellarators for fusion were shown to be invalid once analytic theory developed a coordinate system \cite{Mag-coord}, now known as Boozer coordinates, and simple equations for the drift motion of particles in those coordinates \cite{Boozer:1984}.  This allowed the rapid particle losses to be eliminated by optimization of the externally produced magnetic field \cite{Nuhrenberg-Zille:1986}.  The W7-X stellarator \cite{W7-X,Long puse:2026} has given an empirical demonstration that this transport issue can be resolved.  

The resolution of the issue of particle confinement in stellarators demonstrates the importance of unexpected innovations in analytic theory advanced by computations, which is the second role of theory and computation.

What was demonstrated forty-five years ago \cite{Mag-coord} is that the magnetic field $\vec{B}(\vec{x},t)$ can be written in two simple forms whenever or wherever magnetic surfaces and a plasma equilibrium, $\vec{\nabla}p=\vec{j}\times\vec{B}$, exist:
\begin{eqnarray}
2\pi \vec{B} &=& \vec{\nabla} \psi_t \times \vec{\nabla}\theta + \vec{\nabla} \varphi \times \vec{\nabla}\psi_p(\psi_t,t);  \label{contra-rep}\\
&=& \mu_0 G(\psi_t,t) \vec{\nabla}\theta +  \mu_0 I(\psi_t,t) \vec{\nabla}\theta + \beta_*\vec{\nabla}\psi_t. \hspace{0.2in} \label{cov-rep}
\end{eqnarray}  
The first or contravariant representation, Equation (\ref{contra-rep}) makes the divergence-free nature of $\vec{B}(\vec{x})$ manifest.  The second or covariant form, Equation (\ref{cov-rep}), allows a simple determination of $\vec{\nabla}\times\vec{B}$.  The mathematics of general coordinates is derived in a two-page appendix to \cite{Boozer:RMP}.

Boozer coordinates $\vec{x}(\psi_t,\theta,\varphi,t)$ mean that the ordinary Cartesian coordinates $(x,y,z)$ can be given as functions of $(\psi_t,\theta,\varphi)$ and time $t$, Figure \ref{fig:fluxes-currents}.   Using both representations, the guiding center motion of particles can be obtained in the $(\psi_t,\theta,\varphi)$ coordinates without knowing $\vec{B}(\vec{x})$ only the field strength is required \cite{Boozer:1984}.

Two of the coefficients that appear in Equation (\ref{cov-rep}) for the covariant representation, the poloidal current $G$ and the toroidal current $I$, are  just functions of $\psi_t$ and time.  The more complicated dependence of the coefficient $\beta_*(\psi_t,\theta,\varphi,t)$ creates no fundamental complications.  It is determined by the pressure gradient $dp/d\psi_t$ with $\vec{\nabla}\beta_*\times\vec{\nabla}\psi_t$ giving what is known as the Pfirsch-Schl\"uter current.

Boozer coordinates are required to understand modern stellarator designs, so their implications are well known among those who study stellarators. But, many who study tokamaks have far less familiarity.  The perceived simplicity of axisymmetric tokamaks encourages the development of ad hoc models which may or may not be valid.  Lack of familiarity may make what is profound seem naive and what is naive seem profound.  A clear example is the implications of Faraday's Law, which are discussed in Section \ref{sec:Faraday}.  Confusions about the implications of Faraday's Law led to a claim in a Nuclear Fusion paper \cite{Fitzpatrick:2026} that the conclusions about disruptions given here are incorrect.

Two sections clarify the application of the mathematics associated with Boozer coordinates to tokamaks.  Section (\ref{sec:Faraday}) is focused on Faraday's Law and discusses how ignoring simple but exact results led to the false conclusions of the Nuclear Fusion paper \cite{Fitzpatrick:2026}.

%%%%%

 \color{black} 

Section (\ref{sec:ell_i})  introduces the shape function $\sigma(\psi_t,t)$, which in axisymmetry gives the effect of the shaping of the magnetic surfaces on the safety factor $q(\psi_t,t)$ and on the internal energy inductance $\ell_i$.  The method for doing this was introduced in Section 5.3.4 of a 2015 review \cite{NF review}.  

The expressions using the shape function dispel another confusion in the Nuclear Fusion paper \cite{Fitzpatrick:2026} that the elliptical shaping modifies the effect of the current profile through an empirically determined factor.   The shape function is used in \cite{Boozer:sep2026} to obtain the effect of the separatrix of a divertor on $\ell_i$ and $q(\psi_t)$, which are often used as the primary parameters in tokamak disruption studies.  

\color{black}

%%%%%%%%

When axisymmetry is broken, the magnetic field lines can become chaotic, which means that infinitesimally separated field lines throughout a volume have a separation that depends exponentially on the distance $\ell$ along the lines.  Remarkably, an ideal perturbation can produce a chaotic region by distorting while preserving nested magnetic surfaces.  This explains \cite{Boozer:surf-des}  the rapid destruction of the magnetic surfaces that characterize many disruptions.  The importance to the speed of disruptions was quickly confirmed in simulations of tokamak experiments by Jardin et al \cite{Jardin:2022}.

Even in the absence of magnetic surfaces,  Equation (\ref{contra-rep}) for the contravariant representation of the magnetic field remains valid except $\psi_p$ becomes a function of all three coordinates $(\psi_t,\theta,\varphi)$ as well as time.  In the absence of magnetic surfaces, the covariant representation, Equation (\ref{cov-rep}), is not generally valid except for curl-free magnetic fields, which make $G$ a spatial constant.  The effects of chaos are of central importance to stellarators, tokamaks, and particularly magnetic reconnection, but they are not well known among plasma physicists.  A review, \emph{Magnetic Field Line Chaos, Cantori, and Turnstiles in Toroidal Plasmas} \cite{Boozer:Chaos} has recently been published.  A turbulent magnetic field is chaotic, but a chaotic magnetic field need not be turbulent.    When the magnetic field is turbulent, the rate of dissipation of its helicity by resistivity is not enhanced, unlike its energy.  The Appendix discusses the implications for tokamaks.

ARC and STEP will be discussed as prototype power-plant designs.  Commonwealth Fusion Systems (CFS) is proceeding on a fast timescale to operate the SPARC tokamak \cite{SPARC Overview2020} in order to develop the knowledge needed for a demonstration fusion power plant, ARC \cite{ARC2026}.  It is of importance to clarify operational constraints and control issues of tokamak power plants, such as ARC, and how these issues can be addressed using SPARC. \color{black} This paper is intended to provide some of that clarification. \color{black} The United Kingdom Atomic Energy Authority (UKAEA) is on a similar timescale with its STEP prototype spherical tokamak power plant \cite{STEP:2024,STEP-Overview}.

Extrapolations based on decades of non-ignited tokamak experiments by thousands of people underlie the development of the designs for ARC and STEP.  Two points should be considered.  First, extrapolations do not override the validity of fundamental physics.  For example, Faraday's Law, not large extrapolations from experiments, must be trusted if there are disagreements.   Second, the empirical finding that $n\tau_ET$ rapidly degrades with pulse length must be shown to be misleading, either by experiments or by a convincing theory.  Should this be an objective of SPARC before the design of ARC is settled?
  
Diversity of concepts is important for the development and optimization of fusion.  Nevertheless, the development of fusion with minimal time and cost requires an informed allocation of resources among the various concepts using not only empirical assessments but also theory and computation.

A short summary of the paper is given in Section \ref{Sec:Discussion}.

%%%%%%%%%%%%%%%%%%%%%%%%%%%%%%%%

\section{Faraday's Law \label{sec:Faraday}}  % The greatest danger from ignorance arises when you think you know but do not.

\color{black}

The tokamak community has focused so strongly on the poloidal magnetic flux $\psi_p$ that the toroidal flux $\psi_t$ is rarely mentioned.  But, Faraday's Law gives an exact statement about the slippage of the poloidal relative to the toroidal magnetic flux.  This statement,  Equation (\ref{flux-ev}), has the certainty of Faraday's Law whenever and wherever magnetic surfaces persist that are defined by the toroidal flux $\psi_t$ that they enclose.  For consistency, mathematics requires the poloidal flux be defined as is the flux through the central hole of a toroidal magnetic surface,  Figure \ref{fig:fluxes-currents}.  

The failure to appreciate these points about Faraday's law led to the version of this paper that was submitted to the Physics of Plasmas on September 17, 2025, being flatly rejected on November 10, 2025.  The first reviewer stated: ``The discussion of the plasma current ramp-down is similarly qualitative, and a more rigorous model was recently published in Nuclear Fusion by Fitzpatrick \cite{Fitzpatrick:2026}, prompted by the arXiv article from this author \cite{Constraints:V5}.''  Unfortunately, the Nuclear Fusion paper  \cite{Fitzpatrick:2026} was based on fundamental misunderstandings about Faraday's Law.  But, this neither blocked the exuberance of the reviewer of \cite{Constraints:V5}, nor prevented its acceptance and November 6 online publication by Nuclear Fusion.  Before the Physics of Plasmas would accept a resubmission, the criticisms of the Nuclear Fusion paper had to be addressed, and this was done in \cite{Comment}.  

The failure to use the toroidal flux as the radial coordinate not only makes it difficult to avoid mistakes in determining the evolution of the poloidal flux it also confuses the definition of the edge safety factor in tokamaks with a divertor.  The safety factor on the separatrix of a tokamak divertor is infinite, so the safety factor at the edge is identified with $q_{95}$, which is the safety factor on the magnetic surface that has 95\% of the poloidal flux that lies between the magnetic axis and the separatrix \cite{Def-q_95}.  This definition couples the geometric issues of a separatrix with the central current profile \cite{Boozer:sep2026}.  The use of the toroidal flux decouples the two.  Ninety-five percent of the poloidal flux between the axis and the separatrix corresponds to from 80\% to 90\% of the toroidal flux enclosed by the separatrix.

To be definite, $\psi_t = 0.85 \Psi_t^{sep}$ is chosen as the definition of the plasma edge, where $\Psi_t^{sep}$ is the toroidal flux enclosed by the separatrix.  This definition is denoted for the safety factor and other quantities by $q_{t85}$, but $q_{t85}$ is approximately equal to $q_{95}$.

\color{black}

Before discussing subtleties, Stokes' Theorem will be applied to Faraday's Law at the circular magnetic axis of a tokamak.  This application demonstrates problems with the analysis given in the Nuclear Fusion paper  \cite{Fitzpatrick:2026}.  As is clear from Figure \ref{fig: B}, the magnetic flux $\Psi_p^{ax}(t)$ going down through the circle defined by the axis is the sum of two fluxes:  $\Psi_p^{sol}(t)$, the flux in the central solenoid, and $\Psi_p^{pl}(t)$, the flux produced by the plasma current. 

%%%%%%%%%%%%%%%%%%%%%%%%% 

\subsection{Magnetic axis Faraday's Law result}

Stokes' Theorem applied to Faraday's Law  at the circular magnetic axis implies
\begin{eqnarray}
\frac{d\Psi_p^{ax}}{dt} &=& \oint_{ax} \vec{E}\cdot d\vec{\ell} \\
&\equiv&V_{\ell}^{ax}(t),
\end{eqnarray}
where $d\vec{\ell}$ is the differential distance along the magnetic field line that lies along the axis and $V_{\ell}^{ax}$ is the loop voltage on the axis.  Similarly the loop voltage due to a flux swing in the central solenoid is
\begin{eqnarray}
V_{\ell}^{sol} &=& \frac{d\Psi_p^{sol}}{dt}, \mbox{   so  } \\
V_{\ell}^{ax}&=& \frac{d\Psi_p^{pl}}{dt} + V_\ell^{sol}. \label{V-rel}
\end{eqnarray}
 Although $V_\ell^{sol}(t)$ is the only term in Equation (\ref{V-rel}) that can be directly controlled,  it is never mentioned in the Nuclear Fusion analysis \cite{Fitzpatrick:2026}.    
 
 The magnetic  flux $\Psi_p^{pl}$ due to the toroidal plasma current comes up---not down---through the circle defined by the axis, which means it is negative, Figure \ref{fig: B}.  $\Psi_p^{pl}$ is the sum of a part interior to the plasma, $\Psi_p^{in}$ and a part that is exterior, $\Psi_p^{ex}$.
 \begin{equation}
 \Psi_p^{pl} = - (\Psi_p^{in} +\Psi_p^{ex}).
 \end{equation}

\begin{figure}
\centerline{ \includegraphics[width=2.5 in]{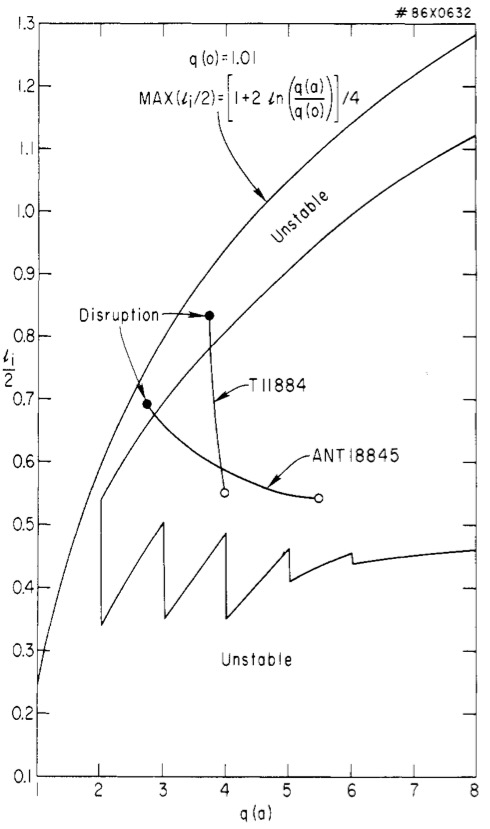}}
\caption{C.Z. Cheng et al, Plasma Phys. Control. Fusion \textbf{29} 351 (1987), studied the regions in $\ell_i$-$q(a)$ space in which tearing-mode stable tokamak current profiles could be obtained.  They found that TFTR plasmas disrupted only when they left the stable region.  They used a cylindrical model with the plasma having a minor radius $a$, a periodicity length $2\pi R$, and a constant ``toroidal" magnetic field $B_t$ along the cylinder.     The degree to which the current profile is highly peaked is quantified by the internal inductance $\ell_i$, Equation (\ref{ell_i cir}), the larger $\ell_i$ the more peaked the current density.  The safety factor $q(a)$ is at the plasma edge. The central safety factor $q(0)$ was assumed to be unity.  Note that the vertical axis is $\ell_i/2$, not $\ell_i$.   }
\label{fig: Disrup}
\end{figure}

In the Nuclear Fusion paper \cite{Fitzpatrick:2026}, the magnetic surfaces were assumed to be circles about an axis of length $2\pi R$ with the plasma edge at a radius $r=a$.  The toroidal flux $\psi_t =B_\varphi \pi r^2$. The poloidal magnetic field $B_\theta(r)=\mu_0I(r)/2\pi r$ gives the poloidal flux due to the plasma current that lies within the plasma,
\begin{equation}
\Psi_p^{in}=  \mu_0 R \int_0^a \frac{I(r)}{r} dr.
\end{equation}
Actually a constant factor $\gamma_s$ for plasma shaping was introduced in the calculations, which is of no fundamental importance.   This factor is inconsistent with the use in the paper  \cite{Fitzpatrick:2026} of the Cheng et al stability diagram, \cite{MHD stab} and Figure \ref{fig: Disrup}, to assess stability against disruptions.  

What was totally missed in the Nuclear Fusion paper \cite{Fitzpatrick:2026} is the poloidal flux produced by the plasma current that lies outside of the plasma $\Psi_p^{ex}$, which would be infinite if the circular surface approximation were continued to $r\rightarrow\infty$.  A finite answer is obtained using  the flux 
\begin{equation}
\Psi_p^{ex}=\Big(\ln\left(\frac{8R}{a}\right)-2\Big) \mu_0R I_p.  \label{ext-flux}
\end{equation}
that comes up through the central hole in a toroidal shell with a major radius of the shell $R>>a$ that encloses a toroidal current $I_p$ in the region $r<a$.  This expression is commonly used and  can be obtained using the method of the solved problem 5.32 in \cite{Jackson}.   

For standard tokamaks, the external part of the poloidal flux produced by the plasma current $\Psi_p^{ex}$ is larger than the flux $\Psi_p^{in}$ given by the plasma current that lies within the plasma.  Equation (21) of the Nuclear Fusion paper \cite{Fitzpatrick:2026} gives a typical internal poloidal flux is $\Psi_{p0}^{in} = 2\pi R a B_\theta(a)$, when the increased current required for a given flux due to non-circularity is ignored.   The ratio  $\Psi_p^{ex}/\Psi_{p0}^{in} = \ln(8R/a) - 2 \approx 1.178$ when $R/a=3$.  For a parabolic current profile, the ratio of the external to the internal flux produced by the plasma current is even larger, $\Psi_p^{ex}/\Psi_p^{in} =1.57$.

Using Cheng et al stability diagram, Figure \ref{fig: Disrup}, to assess stability, only a limited range of current profiles $I(r)/I_p$ are consistent with stability.  As will be discussed in Section \ref{Sec:disruption}, only a moderate change in $\Psi_p^{in}$ occurs over the range of $I(r)/I_p$ consistent with stability.  Consequently, only a small change $\Delta\Psi_p^{pl}/\Psi_p^{pl}$ is consistent with stability against disruptions.

 %%%%%%%%%%%%%%%%%%%%%%
%%%%%%%%%%%%

Assuming the general validity of circular-surface result that $\Delta\Psi_p^{pl}/\Psi_p^{pl}$ must be small compared to unity to avoid disruptions, Equation (\ref{V-rel}) implies
 \begin{eqnarray}
\Big|\frac{V_\ell^{ax} - V_\ell^{sol} }{\frac{d\Psi_p}{dt}}\Big|  \approx \Big|\frac{\Delta\Psi_p}{\Psi_p}\Big| <<1. \label{constancy of voltage}
\end{eqnarray}
This implies the loop voltage is essentially a spatial constant across the plasma but can depend on time.  

The Nuclear Fusion paper \cite{Fitzpatrick:2026} mentioned neither $V_\ell^{sol}(t)$ nor $\Psi_p^{ex}$.  The paper did retain the loop voltage at the axis and implicitly assumed it was controllable by making a strong assumption.  It was assumed that as the magnitude of the poloidal field $B_\theta(r,t)$ evolves it maintains the same profile $B_\theta(r,t)/B_\theta(a,t)$.  This implies the temporal constancy of the current profile $I(r,t)/I_p(t)$ as $I_p$ evolves, which ensures the plasma is stable to disruptions for all points in time when it is initially stable.  The spatial constancy of loop voltage is also implied.  As just shown, the spatial constancy of $V_\ell$ is an accurate approximation.  However, the spatial constancy of the loop voltage does not imply the time independence of the poloidal field profile.

To fully understand why the spatial constancy of the loop voltage does not imply the time independence of the poloidal field profile, the relation between the evolution of the poloidal flux produced by the plasma current $\psi_p(\psi_t,t)$ at each point in the plasma to the local loop voltage must be known, which requires the general Faraday's Law result of Section \ref{Gen-Faraday}.

%%%%%%%%%%%%%%%%%%%%%%%%%%

\subsection{General Faraday's Law result \label{Gen-Faraday}}

The exact relation between the evolution of the poloidal flux and the local loop voltage $V_{\ell}(\psi_t,t)$ may seem obvious:
\begin{eqnarray}
\frac{\partial \psi_p(\psi_t,t)}{\partial t} &=& V_\ell(\psi_t,t); \label{flux-ev} \\
V_\ell(\psi_t,t) &\equiv& \lim_{L\rightarrow\infty} \frac{\int_{-L}^L \vec{E}\cdot d\vec{\ell}}{\int_{-L}^L \vec{\nabla}\left(\frac{\varphi}{2\pi}\right)\cdot d\vec{\ell}},
\end{eqnarray}
where $d\vec{\ell}$ is the differential distance along a magnetic field line in that surface.  But, this result and its generality are actually profound.

The derivation of Equation (\ref{flux-ev}) requires the use of the contravariant representation of the magnetic field, Equation (\ref{contra-rep}), and the theory of general coordinates as was explained in a Reviews of Modern Physics paper \cite{Boozer:RMP} more than twenty years ago.  Wherever and whenever toroidal magnetic surfaces exist, the magnetic field has a simple contravariant representation, Equation (\ref{contra-rep}), and Equation (\ref{flux-ev}) holds for the evolution of $\psi_p(\psi_t,t)$ on each magnetic surface in the plasma with the same certainty as Faraday's Law.  

It is easily shown that the assumption of time independent poloidal field profile, $B_\theta(r,t)/B_\theta(a,t)$, of the Nuclear Fusion paper \cite{Fitzpatrick:2026} is easily violated using Equation (\ref{flux-ev}).  The temporal constancy of $B_\theta(r,t)/B_\theta(a,t)$ with circular surfaces is equivalent to the the temporal constancy of the toroidal current profile $I(r,t)/I_p$.  More generally, the temporal constancy of the toroidal current profile $I(\psi_t,t)/I_p$  does not follow from the spatial constancy of the loop voltage.   

To make the discussion more intuitive, an approximate expression will be used for the density of the net parallel current 
\begin{equation}
\bar{j}_{||}(\psi_t,t)\equiv \bar{B} \frac{\partial I}{\partial\psi_t},
\end{equation}
where $\bar{B}(\psi_t,t)$ is a characteristic magnetic field strength on a particular magnetic surface.  When the loop voltage $V_\ell(\psi_t,t)$ is the same on every magnetic surface, it equals the loop voltage at the magnetic axis, $V_\ell(\psi_t,t)= V_\ell^{ax}(t)$.  The density of the net parallel current is then
\begin{equation}
\bar{j}_{||}(\psi_t,t) = \frac{V_\ell^{ax}(t)}{2\pi R \eta_{||}(Z_{eff},T_e)} + j_{cd} + j_{bs}.  \label{Const-V j profile}
\end{equation}
The dissipation of the current carried by near thermal electrons is given by a resistivity $\eta_{||}$, but that current density can be augmented by external current drive $j_{cd}$ and a bootstrap current density $j_{bs}$.  For reasons discussed in Section \ref{Sec:plasma maintenance}, both $j_{cd}$ and $j_{bs}$ are usually negligible.   They were assumed to be zero in the Nuclear Fusion paper \cite{Fitzpatrick:2026}.  The dominant functional dependencies of $\eta_{||}$ are on the electron temperature $T_e$ and local charge state of the impurities $Z_{eff}$.  The neoclassical corrections to $\eta_{|||}$ are being ignored for simplicity as they were in \cite{Fitzpatrick:2026}.  That paper also assumed that $Z_{eff}$ was a constant in space and time and used an evolution equation that made the electron temperature proflie $T_e(\psi_t,t)/T_e(0,t)$ independent of time.

The time independence of the electron temperature and $Z_{eff}$ profiles, which are required to make the toroidal current profile $I(\psi_t,t)/I_p(t)$ time independent are far from obvious during the shutdown period of a fusion plasma.  During the shutdown, the fusion power quickly terminates leaving only Ohmic heating, which makes a major change in the heating profile.  Impurity radiation is not a negligible power loss and is a very complicated function of the electron temperature even when the impurity content is constant.  At some point the usually assumed H-mode is lost, which also makes a large change in the $T_e$ profile.  In shutdowns in JET-ILW it was found \cite{Sozzi:2020} that carefully controlled heating was required to avoid disruptions.  In a power plant, it is unclear whether sufficient diagnostics or controllable heating would be available.

The author of the Nuclear Fusion paper  \cite{Fitzpatrick:2026} stated in a later arXiv article \cite{Fitzpatrick:2026A}: ``The transport of electron energy in tokamak plasma has been the subject of extensive research for the last 50 years. There is overwhelming evidence that such transport is diffusive in nature.''  Despite the certainty of this statement, a general view among tokamak experts is that energy diffusivity is not always the correct way to characterize energy transport---even when radiative losses are ignored.  The temperature profiles can be given rigidity by microturbulence, \cite{Holland:2021,Turb-profiles:2025}.  The study by N.T. Howard et al of microturbulent transport in ITER \cite{Turb-profiles:2025} has sufficient information to calculate the loop voltage and determine its spatial constancy.   This would determine the consistency of their assumed current profile $I(\psi_t)$.  Much work remains to establish the consistency of current profiles with both thermal transport and stability against disruptions in full ramp-up, flattop, and ramp-down scenarios.   

For tokamaks to be feasible as power plants, a set of plasma parameters need to be identified that can be measured and controlled in a power plant that ensure a sufficiently low disruptivity.  The JET-ILW experiments, \cite{JET-dis:2020} and Figure \ref{fig:JET}, imply the internal inductance and the edge safety factor are not an adequate set.  What will be studied in Section \ref{Sec:disruption} is how the current profile, and in particular current near the plasma edge affects the internal inductance $\ell_i$ for a given edge to central safety-factor ratio.  The pedestal of a tokamak H-mode increases the electron temperature and has a strong bootstrap current, both of which reduce $\ell_i$ for a given safety-factor ratio. \color{black} The geometric effect of the divertor separatrix on the internal inductance for a given edge safety factor is studied in \cite{Boozer:sep2026} through the use of Jacobi elliptic functions to obtain an analytic solution.   These geometric effects are included qualitatively in Section \ref{Sec:disruption}.

%%%%%%%
\color{black}

The importance of exploring a range of relevant current profiles was made manifest by the Nuclear Fusion paper \cite{Fitzpatrick:2026}, which in the end considered only one current profile and used its stability to argue for general stability against disruptions.  

As discussed in Section \ref{Sec:plasma maintenance}, a tokamak is expected to require a few thousand pulses a month uninterrupted by any disruptions that require a shutdown that has major implications on the affordability or reliability of fusion power.  It seems difficult to claim that the single current profile of the ``more rigorous model" of the Nuclear Fusion paper  \cite{Fitzpatrick:2026} obviates the need for understanding how an evolution into a disruptive state occurs.  Section \ref{Sec:disruption} on disruptions shows that Equation (\ref{flux-ev}), which depends only on Faraday's Law and mathematics, gives important insights into that.

%%%%%%%%%%%%%%%%%%%%%%%%%%%%%%%%%%%%%%%%%%%%%%

\section{Effect of magnetic-surface shaping \label{sec:ell_i}}

Expanding on a method developed in Section 5.3.4 of a 2015 review \cite{NF review}, simple but exact expressions for tokamaks can be obtained for the safety factor, $q=1/\iota$, the poloidal flux inside the plasma due to the plasma current, $\Psi_p^{in}$, and the internal inductance, $\ell_i$, which measures the peakedness of the plasma current profile.  These expressions hold for arbitrary shaping of the magnetic surfaces.  

The derivations start with an alternative  expression for the contraviant representation of Equation (\ref{contra-rep}):
\begin{eqnarray}
2\pi \vec{B} &=& \Big( \frac{\partial\vec{x}}{\partial \varphi} + \iota \frac{\partial\vec{x}}{\partial \theta} \Big) \frac{1}{\mathcal{J}}, \label{contra2}
\end{eqnarray}
where $\mathcal{J}$ is the $(\psi_t,\theta,\varphi)$ coordinate Jacobian.  This follows from the theory of general coordinates, which are derived in a two-page appendix to \cite{Boozer:RMP}.

%%%%%%%%%%%%%%%% 

\subsection{Effect on the safety factor $q\equiv1/\iota$}

In a toroidally symmetric magnetic field, as in an ideal tokamak, the dot product of the toroidal $\partial\vec{x}/\partial\varphi$ and the poloidal $\partial\vec{x}/\partial\theta$ tangent vectors is zero.   As explained in the appendix to  \cite{Boozer:RMP}, the tangent vector of a coordinate always gives zero when dotted into a gradients of other coordinates and unity when dotted in the gradient of the same coordinate. 

When the contravariant representation, Equation (\ref{contra2}) and  the covariant representation Equation (\ref{cov-rep}) are first dotted with $\partial\vec{x}/\partial\theta$ and then with $\partial\vec{x}/\partial\varphi$, the result is the equations
\begin{eqnarray}
\iota \Big(\frac{\partial\vec{x}}{\partial \theta}\Big)^2 &=& \mu_0 I \mbox{    and  } \\
\Big(\frac{\partial\vec{x}}{\partial \varphi}\Big)^2 &=& \mu_0G \mbox{    so   } \\
\iota(\psi) &=& \frac{I(\psi_t)}{G(\psi_t)} \frac{(\partial\vec{x}/\partial\varphi)^2}{(\partial\vec{x}/\partial\theta)^2} \\
&=&   \frac{I(\psi_t)}{\sigma(\psi_t)G(\psi_t)}  \mbox{    where  } \label{iota-sigma}\\
\sigma(\psi_t)  &\equiv& \frac{(\partial\vec{x}/\partial\theta)^2}{(\partial\vec{x}/\partial\varphi)^2} 
\end{eqnarray}
is the shape function of the magnetic surfaces.  In axisymmetric systems, the spatial dependence of the shape function is given by $\psi_t$ alone since Equation (\ref{iota-sigma}) relates $\sigma$ to functions of $\psi_t$.  The situation in a stellarator is more complicated because the dot product $(\partial\vec{x}/\partial\varphi)\cdot(\partial\vec{x}/\partial\theta)$ is non-zero and gives the contribution to the rotational transform that is present even in curl-free magnetic fields, Section 5.3.4 of \cite{NF review}. 

When the magnetic surfaces are circular in $(r,\theta,z)$ cylindrical coordinates with periodicity in $z=R\varphi$, where $\varphi$ represents the toroidal angle,
\begin{eqnarray}
\sigma(r) &=& \frac{r^2}{R^2} \mbox{   and   } \label{sigma-circle} \\
\iota(r) &=& \frac{R}{r}\frac{2\pi R}{\mu_0G}\frac{\mu_0 I(r)}{2\pi r }\\
&=& \frac{RB_\theta}{rB_\varphi},
\end{eqnarray}
a well known expression

%%%%%%
\color{black}
An especially simple expression for the safety factor, $q(\psi_t)\equiv1/\iota(\psi_t)$, is obtained when it normalized by its value at the plasma boundary, which encloses a toroidal flux $\Psi_t$.  There is a subtlety when the plasma boundary is a separatrix for $\sigma$ goes to infinity there.  The convention \cite{Def-q_95} is to define the edge as the location at which $\psi_p-\psi_p^{ax} = 0.95 (\psi_p^{sep}-\psi_p^{ax})$, the place where the difference between the poloidal flux and that at the axis equals 95\% of the difference between the poloidal flux at the separatrix and that at the axis.  As noted in Section \ref{sec:Faraday} this convention needlessly requires knowledge of the current density in the central part of the plasma in order to define the plasma edge.  

To avoid the complications that follow from this unfortunate convention of defining the edge using a poloidal rather than a toroidal flux fraction, this paper will define the plasma edge as $\psi_t=\Psi_{t85}$, which is where approximately  95\% the poloidal flux difference is reached \cite{Boozer:sep2026}.   The simple expression for the safety factor is then
\begin{equation}
\frac{q(\psi_t)}{q_{t85}} = \frac{I_{t85}}{I(\psi_t)} \frac{\sigma(\psi_t)}{\sigma_{t85}}. \label{q profile}
\end{equation}
Given the importance of determining parameters that ensure low disruptivity, studies similar to those in \cite{Def-q_95} should be done to determine the optimal choice of the toroidal flux fraction; 85\% seems low.

Although the current $I_{t85}$ approximates the plasma current $I_p$, the two are different in an H-mode plasma, which has a strong current density in the thin pedestal region at the plasma edge.   This effect can be represented by a delta-function current density 
\begin{equation}
\left(\frac{dI}{d\psi_t}\right)_{edge} = I_e\delta(\psi_t - \Psi_{t85}). \label{I-edge}
\end{equation}
$I_e$ is the edge or pedestal current, which makes $I_{t85}$ equal the plasma current $I_p$.

%%%%%%%%%%%%%%%%%%%%%

Equations and their interpretations become simpler by defining
\begin{equation}
s\equiv \frac{\psi_t}{\Psi_{t85}},
\end{equation}
and taking $s=1$ to be what is meant by the plasma edge whether there is a separatrix or not. 

  %%%

 The simplest approximation to the effect of a separatrix \cite{Boozer:sep2026} is to choose a constant $c_s$ such that
 \begin{eqnarray}
\frac{\sigma(s)}{\sigma_{t85}} & \approx& \frac{s + c_s s^2}{1+c_s} \mbox{    so } \label{sigma-sep}\\
  \frac{q(s)}{q_{t85}} &=& \frac{I_{t85}}{I(s)} \frac{\sigma(s)}{\sigma(s=1)} \\
 & \approx&\frac{I_{t85}}{I(s)} \frac{s+ c_s s^2}{1+c_s}  \mbox{    with } \label{simple-q} \\
c_s & \approx& 0.32
 \end{eqnarray}
 with a separatrix and zero without a separatrix.   The value of $c_s=0.32$ corresponds to $\Pi_\sigma=0.38$ at $\psi_t/\Psi_t =0.85$ in \cite{Boozer:sep2026}.
 
 As $s\rightarrow0$
 \begin{eqnarray}
 \frac{q(0)}{q_{t85}} &=& \frac{I_{t85}}{\left(\frac{dI}{ds}\right)_0} \frac{1}{1+c_s} \mbox{  or   } \\
 \frac{\left(\frac{dI}{ds}\right)_0}{I_{t85}} &=& \frac{q_{t85}}{(1+c_s)q(0)}, \label{cental current}
 \end{eqnarray}
 where $(dI/ds)_0 = \Psi_{t85} (dI/d\psi_t)_{ax}$ with the value of $j_{||}/B$ along the magnetic axis equal to $(dI/d\psi_t)_{ax}$.
 
 %%%%%%%%%%%%%%%%%%%

%%%%%%%%%%%%%%%%%%%%%%%%%%

\subsection{Effect on $\Psi_p^{in}$ and $\Lambda_{in}$}

%%%%%%%

The poloidal magnetic flux produced by plasma current that lies within the plasma is
\begin{eqnarray}
\Psi_p^{in} &=& \int_0^{\Psi_t} \iota(\psi_t) d\psi_t \\
&=& \int_0^{\Psi_t}    \frac{I(\psi_t)}{\sigma(\psi_t)G(\psi_t)} d\psi_t.
\end{eqnarray}

Although this integral is well defined despite the singularity of $\sigma(\psi_t)$ on the separatrix, the expression for the poloidal flux within the plasma edge, which is defined by $\Psi_{t85}$, is also useful.
\begin{eqnarray}
\Psi_p^{t85}&=& \frac{2\Psi_{t85} I_{95}}{\sigma_{t85} G_{t85}} \Lambda_{in} \hspace{0.2in} \mbox{   where   } \\
\Lambda_{in} &\equiv& \frac{1}{2} \int_0^1  \frac{I(s)}{I_{t85}}\frac{\sigma_{t85}}{\sigma(s)} \frac{G_{t85}}{G(s)}ds \label{Lambda-exp}
\end{eqnarray}
is a normalized flux inductance, which gives the effect of the current profile on $\Psi_p^{in}$.  

For the circular cylindrical model of the magnetic surfaces in which $2\pi R$ is the distance of a period along $z$, the toroidal flux is $\Psi_{t95}=\pi B_\varphi a^2$ with $B_\varphi =\mu_0G/2\pi R$, and the shape function is $a^2/R^2$,
\begin{eqnarray}
 \frac{2\Psi_{t85} }{\sigma_{t85} G_{t85}} &=&\mu_0 R
 \end{eqnarray}
 
 \color{black} %%%%%%%%%%%%%%%%%%%%%%%
 
 The definition of $\Lambda_{ex}$ that is consistent with the definition of $\Lambda_{in}$ of Equation (\ref{Lambda-exp}) is 
  \begin{eqnarray} 
  \Lambda_{ex} &=& \ln\left(\frac{8R}{a}\right)-2 \label{Lambda_ex} \\ 
  &=& 1.178 \end{eqnarray} 
  for an aspect $R/a=3$

%%%%%%%%%%%%%

\subsection{Effect on the internal inductance $\ell_i$}

The internal inductance for the plasma energy is a dimensionless measure of the peakedness of the plasma current $I(\psi_t)$ and appears in expressions for plasma stability and for the Shafranov shift.

A general representation of the energy density of the magnetic field in the presence of magnetic surfaces is
\begin{eqnarray}
\frac{B^2}{2\mu_0} &=& \frac{ \left(\vec{B}\cdot\frac{\partial\vec{x}}{\partial \theta}\right) \left(\vec{B}\cdot\vec{\nabla}\theta\right)  + \left(\vec{B}\cdot\frac{\partial\vec{x}}{\partial \varphi}\right) \left(\vec{B}\cdot\vec{\nabla}\varphi\right) }{2\mu_0}. \hspace{0.21in}
\end{eqnarray}
The first term is the energy density of the poloidal and the second the toroidal field.  The energy density of the poloidal field $B_p^2/2\mu_0$ can be obtained by using Equation (\ref{cov-rep}) to calculate $\vec{B}\cdot \left(\partial\vec{x}/\partial\theta)\right)$ and Equation (\ref{contra2}) to calculate $\vec{B}\cdot\vec{\nabla}\theta$:
\begin{eqnarray}
\frac{B_p^2}{2\mu_0} &=& \frac{I(\psi_t) \iota(\psi_t)}{2(2\pi)^2 \mathcal{J}}, \mbox{   so  } \\
\int \frac{B_p^2}{2\mu_0} d^3x &=&  \int_{pl} \frac{I(\psi_t) \iota(\psi_t)}{2(2\pi)^2 \mathcal{J}} \mathcal{J} d\psi_t d\theta d\varphi\\
&=& \int_0^{\Psi_t} \frac{I(\psi_t) \iota(\psi_t)}{2} d\psi_t \label{int-energy}
\end{eqnarray}
is the poloidal field energy within the plasma, where $\Psi_t$ is the toroidal flux within the boundary.  

\color{black}
The internal inductance of the plasma for energy is defined by normalizing the poloidal field energy in the plasma using values at the plasma edge.   Although the poloidal field energy within the plasma is well defined, what is meant by edge values is not.  Given the simple form of Equation (\ref{int-energy}) for the poloidal field energy in the plasma, a natural and precise definition is 
\begin{eqnarray}
\ell_i &\equiv& \frac {\int_0^1 I(s) \iota(s) ds}{I_{t85} \iota_{t85} }.
\end{eqnarray}
This expression is valid for both tokamaks and stellarators.  For tokamaks, the expression of Equation (\ref{iota-sigma}) for $\iota(\psi_t)$ can be used, which implies an even simper definition
\begin{eqnarray}
\ell_i&\equiv& \int_0^1 \left(\frac{I(s)}{I_{t85}}\right)^2 \frac{\sigma_{t85}} {\sigma(s)} \frac{G_{t5}} {G(s)}ds. \label{ell_i-exp}
\end{eqnarray}

\color{black}%%%%%%%%

When the magnetic surfaces are circular or have a radially constant ellipticity $\sigma(s)/\sigma_{t85}=r^2/a^2$, $ds = 2rdr/a^2$, and $B_\theta(r) = \mu_0I(r)/2\pi r$, so
\begin{equation}
\ell_i = 2 \int_0^a \frac{B_\theta^2(r) rdr}{B_\theta^2(a)a^2}, \label{ell_i cir}
\end{equation}
which is the universal expression for the internal inductance of an axisymmetric tokamak in the circular-surface approximation.

%%%%%%%%%%%%%%%%%%%%%%%%%%%%%%%%%%%%%%%

\section{Disruptions \label{Sec:disruption}}

Tokamak disruptions are a sudden loss of plasma confinement and are a major challenge to achieving tokamak power plants \cite{Eiditis:2021}.  Disruptions are dangerous because they can damage internal components---especially components embrittled by fusion neutrons---due to large force and heat loads and a tendency to transfer the plasma current from thermal to relativistic electrons.  The relativistic electrons can cause extreme damage since they can strike the walls in highly localized beams in space and time \cite{Breizman:2019}.

The tokamak community became convinced of the importance of tearing mode instabilities to disruptions in 1978 after the publication of a Bruce Waddell led study \cite{Waddell:1978}.  

In 1987, Cheng, Furth and Boozer \cite{MHD stab} used a model in which the magnetic surfaces were circular to determine which current profiles are consistent with disruption avoidance and showed their results were consistent with the disruptivity of slowly evolving TFTR plasmas. \color{black}  Their diagrams, one is shown in Figure \ref{fig: Disrup}, depend on two relatively easily measured properties, the safety factor at the plasma edge $q_a$ and the internal inductance for energy $\ell_i$. \color{black}   At a fixed edge safety factor, $q_a$ the upper $\ell_i$ limit is given by tearing modes and the lower limit by kinks.  

The circular-surface model is only heuristic, and a similar study should be done for tokamak designs for power plants.  This has not been done for two reasons.  The relationship between the net-parallel current in the plasma $I(\psi_t)$ and the edge safety factor and $\ell_i$ it produces is more complicated, but this can be determined as was demonstrated in Section \ref{sec:ell_i}.  More importantly, neither three dimensional magnetohydrodynamic evolution codes nor even linear ideal and resistive stability codes have been used at the level required to determine which $I(\psi_t)$ profiles are disruptive and which are not in analogy to the Cheng et al \cite{MHD stab} study.  

The feasibility of tokamak power plants requires that parameters be identified that can be measured and controlled that ensure that disruptions are neither too common nor destructive.  Tokamak experiments as well as theory are required. 

The importance of $\ell_i$ and the edge safety factor for disruptions in the DIII-D tokamak was observed  by Sweeney et al \cite{Sweeney:2017}:  ``the plasma internal inductance $\ell_i$ divided by the safety factor at 95\% of the poloidal flux, $q_{95}$, is found to exhibit predictive capability over whether a locked mode will cause a disruption or not."  Their condition resembles the upper limit on $\ell_i/2$ found by Cheng et al,  \color{black} but Sweeney et al  saw disruptions at significantly lower values of $\ell_i/q_{95}$ than expected from Cheng et al.

\begin{figure}
\centerline{ \includegraphics[width=3.0 in]{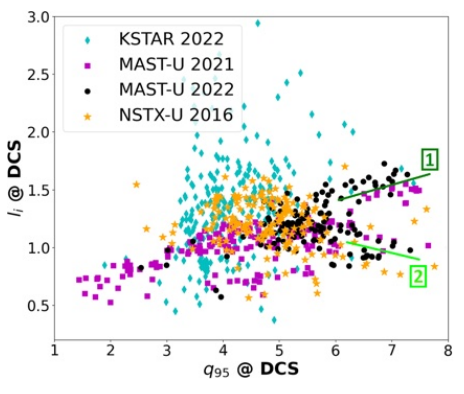}}
\caption{MHD-caused disruptions are illustrated in an internal inductance $\ell_i$ and edge safety factor $q_{95}$ diagram with separate symbols for KSTAR and NSTX-U as well as for two operational years of MAST-U.  Two branches of DCS trigger points can be recognized in both MAST-U experimental years. The two branches (visualized with green lines) are separated along the $\ell_i$ axis into two regions, with the internal inductance being roughly restricted to $1.5 \lesssim \ell_i \lesssim 1.7$  in branch 1 and $0.8 \lesssim \ell_i \lesssim 1.2$ in branch 2 . The region between the branches is free of MHD-caused disruptions for $q_{95}\gtrsim 6.3.$  This is Figure 7 in the paper V. Zamkovska, S.A. Sabbagh, M. Tobin, et al, Nucl. Fusion \textbf{64}, 066030 (2024). }
\label{fig:DCAF_f}
\end{figure}

Steve Sabbagh's group has used the DCAF code \cite{Sabbagh:2024} to determine  the empirical dependence of tokamak disruptions on plasma quantities such as the internal inductance $\ell_i$ and edge safety factor $q_{95}$, Figure \ref{fig:DCAF_f},   The results for MAST-U, which has highly non-circular magnetic surfaces, gave preliminary evidence that an internal inductance that is either too low or too high gives disruptive plasmas with a stable region between, but this split occurs for very large values of the safety factor, $q_{95}>6.5$. 

\begin{figure}
\centerline{ \includegraphics[width=3.0 in]{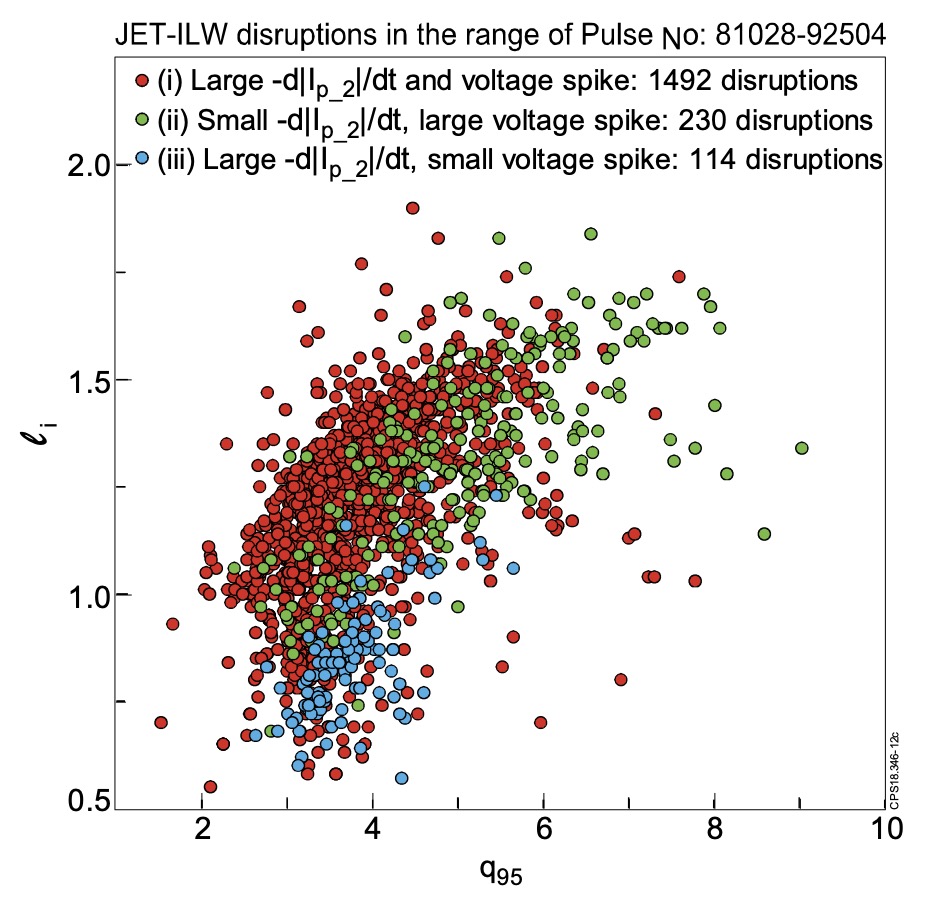}}
\caption{Pre-disruptive parameters in JET with an ITER-like wall are shown in a $\ell_i-q_{95}$ stability diagram.  This was Figure 13 in Gerasimov et al, Nucl. Fusion \textbf{60},  066028 (2020).  As the authors note:  ``It may be expected that a disruption free space may be defined in the $\ell_i-q_{95}$ empirical stability diagram, assuming that plasma current profiles tend to maintain itself inside the permissible values. In reality, the JET-ILW pre-disruptive plasma equilibrium parameters create a diffused cloud on the $\ell_i-q_{95}$ stability diagram without room for non-disruptive plasmas."   }
\label{fig:JET}
\end{figure}

Experiments on the highly-shaped JET tokamak with an ITER-Like Wall (JET-ILW) found the internal inductance $\ell_i$ and $q_{95}$ were not adequate for defining regions of low disruptivity,  Figure \ref{fig:JET}.  They found \cite{JET-dis:2020} that ``there is no way to completely avoid disruptions even for near identical primitive plasma pulses $\cdots$ because there are some uncontrolled causes which lead to disruptions."  Nevertheless, a sufficiently large internal inductance for a given  $q_{95}$ essentially ensures a disruption just as in  the TFTR study.  

The JET-ILW results challenge simple views of disruption control.  They indicate that parameters have not yet been identified that can be measured and controlled that ensure that disruptions are neither too common nor destructive.  Analytic theory, simulations, and experiments can contribute to that identification.

 %%%%%%%%%%%%%%%%%%%%%%
%%%%%%%%%%%% 

%%%%%%%%%%%%%%%%%%%%%%
%%%%%%%%%%%% 

Many reviews have been written on the causes of tokamak disruptions, but they ignore a fundamental reason why disruptions are so common.  Only a small fractional change, approximately $1/6$, in the total poloidal flux $\Psi_p$ produced by the plasma current is sufficient to carry all the way across the stability window of Cheng et al.  As shown in Equation (\ref{constancy of voltage}), this implies the near spatial constancy of the loop voltage across the plasma, which determines current profile, Equation (\ref{Const-V j profile}).

Rather than solving Equation (\ref{Const-V j profile}) for the profile of $I(\psi_t)$, which has many uncertainties, it is better to study which $I(\psi_t)/I_{t85}$ profiles are disruptive and which are not.  This can be easily done with the use of the Cheng et al diagram of Figure \ref{fig: Disrup} to determine whether a particular $I(\psi_t)/I_{t85}$ profile is disruptive.  As crude as this approximation maybe, it does illustrate important points about the robustness of tokamaks to disruptions.  More realistic treatments for tokamak power-plant designs could and should be made but require the resources associated with design groups, not those available to an individual university researcher whose funding for tokamak research has been terminated.

%%%%%%%%%%%%%%%%%%%%%%%%%%%%%%%%%%%%

\subsection{Three-constant current profile \label{3-const}}

A simple current profile, which depends on three constants, can be used to study the effects of $I(s)/I_{t85}$ in the presence or absence of a separatrix. Analytic expressions will be determined for the safety factor ratio, $q_{t85}/q_0$, the internal inductance for energy, $\ell_i$, and $\Lambda_{in}$, the current profile dependent part of the poloidal flux produced by the plasma current that lies within the plasma.  

\subsubsection{Definition of the three-constant current profile}

The three-constant current profile is
\begin{eqnarray}
&\frac{dI}{ds} = I'_0 \hspace{0.2in} &\mbox{for   } s_{st}>s \\
&\hspace{0.2in} = I'_r \hspace{0.2in} &\mbox{for   } 1 > s >s_{st} \hspace{0.2in}\\
&\hspace{0.7in} = I_e\delta(s-1) \hspace{0.2in} &\mbox{at   } s=1, 
\end{eqnarray}
where $s_{st}$ gives the radial extent of the flattening of the current profile by the sawtooth effect, which makes $q_0\equiv q(0)=1$.  The normalized current density in the plasma center is $I'_0/I_{t85}$, the normalized current density in the annular ring is $I'_r/I_{t85}$, and the normalized delta-function current at the plasma edge is $I_e/I_{t85}$.  The poloidial current profile, $G_{t85}/G(s)$ will be assumed to be unity.

The total plasma current $I_{t85}$, $s_{st}$, and the current profile $I(s)/I_{t85}$ are determined by the three parameters $I'_0/I_{t85}$, $ I'_r/I_{t85}$, and $ I_e/I_{t85}$. 
\begin{eqnarray}
&I_{t85} = I'_0 s_s + I'_r (1-s_s) + I_e   \hspace{0.01in} &\mbox{so   } \\
&s_{st} = \frac{1 - \left(\frac{I'_r}{I_{t85}} + \frac{I_e}{I_{t85}}\right)} {\frac{I'_0}{I_{t85}} - \frac{I'_r}{I_{t85}} } \label{s_st} \\
&\frac{I(s)}{I_{t85}} = \frac{I'_0}{I_{t85}} s \hspace{0.01in} &\mbox{for   } s_{st}>s \label{s less s_st}\\
 &\frac{I(s)}{I_{t85}}= \frac{I'_0}{I_{t85}} s_{st} + \frac{I'_r}{I_{t85}} (s-s_{st})  &1>s>s_{st} \hspace{0.1in}  \\
& \frac{I(s)}{I_{t85}} = \frac{I'_0s_{st} }{I_{t85}} + \frac{I'_r(1-s_{st})}{I_{t85}} +\frac{I_e}{I_{t85}}  \hspace{0.01in} &\mbox{for   } s=1. \hspace{0.3in} \label{s of 1}
\end{eqnarray}

\color{black}

The shape function will be studied for two values of $c_s$:  without a separatrix $c_s=0$  and  with a separatrix $c_s=0.32$.  Equation (\ref{sigma-sep}) and (\ref{cental current}) give
\begin{eqnarray}
 \frac{\sigma(s)}{\sigma(1)} &\approx&  \frac{s + c_s s^2}{1+c_s}  \mbox{   and  }\\
 \frac{I'(0)}{I_{t85}} &\approx& \frac{q_{t85}}{(1+c_s)q(0)}. 
 \end{eqnarray}

 The integrals that will be required are
 \begin{eqnarray}
 \int\frac{ds}{s+c_ss^2} &=& \ln\Big(\frac{s}{1+c_ss}\Big) \\
 \int\frac{sds}{s+c_ss^2} &=& \frac{\ln(1+c_s s)}{c_s} \nonumber\\&=& s \mbox{   for   } c_s=0\\
\int\frac{s^2ds}{s+c_ss^2} &=& \frac{s}{c_s}- \frac{\ln(1+c_s s)}{c_s^2} \nonumber\\&=& \frac{s^2}{2}  \mbox{   for   } c_s=0.
 \end{eqnarray}
 
 %%%%%%%%%%%%%%%%
 
 \subsubsection{Expression for the normalized flux inductance }

 The effect of the current profile on the poloidal flux produced by the plasma current that is internal to the plasma is given by the normalized flux inductance $\Lambda_{in}$, Equation (\ref{Lambda-exp}).  
 
 For purposes of comparison, the plasma-produced poloidal flux that comes up through the central hole of the torus defined by the outermost magnetic surface will be taken to be independent of the current profile and given by Equation (\ref{Lambda_ex}) with an aspect ratio $R/a=3$.  That is the coefficient for $\Psi_P^{pl}$ is $\Lambda_{tot} =\Lambda_{in} + \Lambda_{ex}$ with $\Lambda_{ex}=1.178$, and
 \begin{eqnarray}
\Lambda_{in} &\equiv& \frac{1}{2} \int_0^1  \frac{I(s)}{I_{t85}}\frac{ds}{s+s^2} \\
&=& \frac{1}{2}\Big\{\frac{I'_0}{I_{t85}} \frac{\ln(1+c_ss_{st})}{c_s}  \nonumber\\ &&  - \Big( \frac{I'_0}{I_{t85})} - \frac{I'_r}{I_{t85}} \Big)s_{st}\ln\left(\frac{(1+c_s)s_{st}}{1+c_ss_{st}}\right) \nonumber\\ &&+ \frac{I'_r}{I_{t85}}\ln\left(\frac{1+c_s}{1+c_s s_{st}}\right) \Big\}.
\end{eqnarray}
 
The stability of only the case without a separatrix can be correctly assessed by the Cheng et al diagram of Figure \ref{fig: Disrup}.  Nevertheless, the stable range for a given $q_a$ from this diagram will be used to qualitatively assess 
 \begin{equation}
 \frac{\Lambda_{tot}}{\Delta \Lambda} =\frac{1}{2} \frac{(\Lambda_{tot})_{max} + (\Lambda_{tot})_{min}}{\Lambda_{tot})_{max} - (\Lambda_{tot})_{min}} \label{Lambda_tot}
 \end{equation}
 over the region of disruption stability.  
 
 %%%%%%%%%%%%%%%%%%%%
 
 \subsubsection{Expression for the internal inductance $\ell_i$}
 
 The most complicated part of the integral for $\ell_i$ requires $((I(s)/I_{t85})^2$ in the region $1>s>s_{st}$, which is
\begin{eqnarray}
&& \left(\frac{I(s)}{I_{t85}}\right)^2 = \left(\frac{I'_0}{I_{t85}} - \frac{I'_r}{I_{t85}}\right)^2 s_s^2 \nonumber\\&& \hspace{0.2in} + 2\frac{I'_r}{I_{t85}}\left(\frac{I'_0}{I_{t85}} - \frac{I'_r}{I_{t85}}\right)s_{st} s   +  \left(\frac{I'_r}{I_{95}}\right)^2 s^2.\hspace{0.2in} 
\end{eqnarray}

Equation (\ref{ell_i-exp}) gives the integral for the internal inductance,
\begin{eqnarray}
\ell_i &=& \int_0^1 \left(\frac{I(s)}{I_{t85}}\right)^2 \frac{1+c_s} {s+c_ss^2} ds \\
&=& (1+c_s)\Big\{ \left(\frac{I'_0}{I_{t85}}\right)^2\frac{c_s s_{st} -\ln(1+c_s s_{st})}{c_s^2} \nonumber\\
&& +  \left(\frac{I'_0}{I_{t85}} - \frac{I'_r}{I_{t85}}\right)^2 s_{st}^2 \ln\Big(\frac{1+c_s s_{st}}{(1+c_s) s_{st}}\Big)  \nonumber\\&& 
 + 2\frac{I'_r}{I_{95}}\left(\frac{I'_0}{I_{t85}} - \frac{I'_r}{I_{t85}}\right) s_{st}  \ln\Big(\frac{1+c_s}{1+c_s s_{st}}\Big) \nonumber\\&&
  + \left(\frac{I'_r}{I_{t85}}\right)^2 \frac{c_s(1-s_{st}) + \ln\Big(\frac{1+c_s s_{st}}{1+c_s}\Big)}{c_s^2}\Big\}. \hspace{0.2in}
\end{eqnarray}

%%%%%%%%%%%%%%

\subsubsection{Possible stable current profiles}

A simple spread-sheet calculation allows an exploration of the freedom in current profiles that are stable against disruptions.  What is remarkable is that over the full range of the poloidal flux produced by currents in the plasma that are thought to be stable there is only a small fractional change in the normalized flux inductance.  It is the evolution of the poloidal flux, which determines the current profile, so the weak effect that different current profiles have on the poloidal flux makes the avoidance of disruptive current profiles a delicate control problem. 

The plasma flux inductance $\Lambda_{tot}(\ell_i)$ can be determined as a function of the peakedness of the current profile, which means the internal inductance $\ell_i$, for a given edge to central safety factor ratio, with and without a separatrix, and with and without an edge current. The internal inductances $\ell_i$ that will be emphasized are the stability boundaries the Cheng et al diagram of Figure \ref{fig: Disrup} although these are only reliable without a separatrix.   

When the central safety factor is unity and the edge safety factor is $q_{95}\approx q_{t85} =3.5$, tearing modes are stable when $\ell_i<1.52$ and kinks are stable when $\ell_i > 0.8$.  Without a separatrix or edge current, the $\Lambda_{tot}(1.52) = 2.24$ and $\Lambda_{tot}(0.8) = 1.96$.  The average of the two $\Lambda_{tot}$'s divided by their variation is 7.3.  When the plasma is bounded by a separatrix, $\Lambda_{tot}(1.52)=2.02$ and $\Lambda_{tot}(0.8) =1.78$, which gives an average divided by the variation of 7.8.  The presence of separatrix makes a more centrally peaked current profile necessary to achieve a given $\ell_i$.  This is primarily measured by the width of the central region of constant current density, which is the sawtooth width $s_{st}$.  A larger $s_{st}$ gives a more peaked profile.  An internal inductance of $\ell_i=1.52$ requires  $s_{st}=0.26$ without a separatrix but $s_{st}=0.35$ with a separatrix.  A similar effect is seen for $\ell_i=0.8$, without a separatrix $s_{st}=0.18$ and 0.20 with a separatrix.  An edge current of 5\% of the total may realistic but is too small to produce significant changes.

Qualitatively similar results are obtained when the central safety factor is unity and the edge safety factor is $q_{95}\approx q_{t85} =5.5$, tearing modes are stable when $\ell_i<1.9$ and kinks are stable when $\ell_i > 0.9$.  Without a separatrix or edge current, the $\Lambda_{tot}(1.9) = 2.45$ and $\Lambda_{tot}(0.9) = 2.06$.  The average of the two $\Lambda_{tot}$'s divided by their variation is 5.8.  The $s_{st}$ is 0.17 for the larger $\ell_i$ and 0.10 for the smaller.  When the plasma is bounded by a separatrix, $\Lambda_{tot}(1.9) = 2.20$ and $\Lambda_{tot}(0.9) = 1.87$, which gives an average divided by the variation of 6.1.  The $s_{st}$ is 0.21 for the larger $\ell_i$ and 0.11 for the smaller. 

For all the configurations discussed, the ratio of the largest to smallest $\Lambda_{tot} \approx 1.4$, which is smaller than the ratio for $\ell_i\approx 2.4$, or the ratio of $q$'s$\approx1.6$.  What is more important is that all the ratios of $\Lambda_{tot}/\Delta\Lambda_{tot}$ are all large, varying from 5.8 to 7.8.

%%%%%%%%%%%%%%%%%

\subsubsection{Summary of three-constant results}

%====================================

\color{black}

The poloidal flux that must pass through the plasma is enormous compared to the change in the poloidal flux due a change in the current profile that would cross the full range of current profiles that are stable to disruptions.  This is true of startup, shutdown, and a flattop in which the central solenoid can supply poloidal flux comparable to the poloidal flux produced by the plasma current in the period between startup and shutdown.  

The poloidal flux that passes through a plasma can be arbitrarily large compared to the change in the poloidal flux produced by the plasma current only when the loop voltage $V_{\ell}$ is independent of $\psi_t$ for then there is an exact balance between the poloidal flux dissipated at the magnetic axis and that provided by the solenoid and the change in $\Psi_p^{ex}$ as the plasma current  changes.

%%%%%%%%%%%%%

\subsection{Disruption avoidance during tokamak pulses \label{Sec: avoidance}}

Unlike stellarators, tokamaks have passive stability neither for tearing and kink instabilities nor for the axisymmetric plasma position within the chamber.  This makes tokamaks sensitive to unexpected events during the flattop period of plasma pulses---even vertical position control generally requires active feedback.  Oak Nelson et al \cite{SPARC-vert:2024} studied the vertical stability SPARC and included the effect of small disturbances.  

As pointed out in Section \ref{3-const}, tokamaks are sensitive to small deviations in the radial dependence of the loop voltage, which can cause disruptions on a short timescale compared to the characteristic length of plasma pulses, 
 \begin{equation} \tau_{max}\approx\frac{\Psi_p}{V_\ell^{ax}},\end{equation}
that can be maintained by the central solenoid.  A more general expression for $\tau_{max}$ is given in Equation (\ref{tau_max}).    Designs for the central solenoid generally accommodate a flux swing that is only a couple of times larger than the planned plasma-produced poloidal flux $\Psi_p$.

Disruption mitigation is no more an alternative to having methods to ensure disruptive current profiles do not arise than airbags and seat belts allow one to dispense with a steering wheel and brakes in a car.

To avoid disruptions, the current profile could be externally controlled by direct drive of the current or by direct heating to control the profile of the electron temperature.  As discussed in Section \ref{Sec:plasma maintenance}, both require a power comparable to the $\alpha$-heating power in a burning fusion plasma, which makes it difficult to obtain control while maintaining consistency with an energy multiplication factor $Q\sim20$ that is thought to be necessary for the economic feasibility of fusion.

 The shortness of particle confinement time relative to characteristic time for a tokamak pulse, $\tau_{max}$, implies that the particle replacement method could in principle be used to control the temperature and density profiles and thereby the current profile. This presumably requires deep pellet injection.  Subtleties of pellet injection coupled with the subtleties of plasma transport raise many questions.  

The credibility of using particle injection and the adequacy of its control with the available diagnostics is far from assured.  This could be clarified by theory and computations, which could also suggest the design of experiments that could be performed on existing or planned devices such as SPARC, ARC, and STEP.

 Active control requires plasma diagnostics guide the use of actuators to restore the required conditions. In principle, Artificial Intelligence (AI) makes an extremely fast response possible.  Indeed, this was demonstrated \cite{Koleman:2024} for the prevention of disruptions caused by neoclassical tearing modes on DIII-D.   Neoclassical tearing modes \cite{La Haye:2006} are caused by the strong bootstrap current in tokamaks, especially in tokamaks which can maintain a fusion plasma longer than $\tau_{max}$.

The successful application of AI for the active control of neoclassical tearing modes \cite{Koleman:2024} in DIII-D illustrates the difficulty of its application to tokamak power plants.  The DIII-D application relied on continuous profiles of the electron density, electron temperature, ion rotation, safety factor, and plasma pressure to control the neutral beam injection and the triangularity of the plasma shape.  

In a power plant far fewer diagnostics will be available and detailed knowledge of these profiles may not be possible.   The plasma shape is controllable on the timescale on which the currents in superconducting coils can be changed, which is usually several times longer than the penetration time through the chamber walls.  However, the large ports required for neutral beams probably eliminate their use in power plants.  In addition, the injected power must be restricted in order to have sufficient power to sell.

%%%%%%%%%%%%%%%

\subsection{Disruption avoidance during a tokamak shutdown \label{Sec: shutdown}}

The period of plasma shutdown for any large tokamak is an especially difficult period for disruption and runaway-electron avoidance.  Remarkably little has been written on the shutdown of large tokamaks.  P. D. de Vries et al estimated it would take a minute to shutdown ITER \cite{Tok-shutdown:2018}.  As noted by Boozer \cite{Boozer:steering}, the de Vries et al paper did not explain how the plasma current profile could be controlled to avoid disruptions during a one-minute shutdown.     The Nuclear Fusion paper \cite{Fitzpatrick:2026} has given an even more optimistic estimate for how quickly ITER could be shutdown without disruptions and runaway electrons, 14.7 seconds.  As discussed in Section \ref{sec:Faraday}, this analysis has obvious deficiencies \cite{Comment} such as considering only one current profile.  

Plasma disruptions when the plasma current is greater than a few megaamperes are often considered to be too dangerous to allow in a power plant.  Even when a disruption a month is viewed as tolerable, that implies a disruption rate less than one in a thousand pulses.  The issue is not the fastest conceivable time for plasma shutdown but the time required to repeatedly startup, operate, and shutdown the plasma with disruptions sufficiently rare and weak for a practical fusion power plant.  This requires that energy input for control be less than 5\% of the fusion energy output

 To quickly remove the poloidal flux by its dissipation at the axis, the plasma must be cooled, which takes a minimum time $\tau_{min}\approx\left(1+2/(3\beta_p)\right)\tau_E$, with $\beta_p$ the poloidal beta of the plasma and $\tau_E$ the energy confinement time.  The time $\tau_{min}$ is comparable to the shutdown time given in the Nuclear Fusion paper \cite{Fitzpatrick:2026}.  Without active control of the current profile, having less than one in a thousand pulses disrupt, or whatever the required number may be,  seems far from obvious.  
 
 The current profile during shutdown could be controlled by plasma heating and current drive or to a certain extent by maintaining a fixed edge safety factor as suggested in the Nuclear Fusion paper \cite{Fitzpatrick:2026}.  However, the results of Cheng et al \cite{MHD stab} suggest that increasing the edge $q$ increases the range of stable profiles.  Plasma heating and current drive take energy, which can only be a small fraction of the fusion energy produced during the pulse.  Maintenance of an edge safety factor requires a loop voltage, so some fraction of the poloidal flux swing of the solenoid would need to be reserved for this purpose.  Careful simulations of the plasma from startup to shutdown could determine the required fractions of the fusion energy and the solenoidal flux.
 
 In principle, the time derivative of the solenoidal flux $\Psi_{sol}(t)$ can be reversed, which removes flux rapidly from the plasma by reversing the direction of the plasma current near the edge.  Although this strategy does not seem to have extensively studied, the results of  Cheng et al \cite{MHD stab} make one suspicious that the plasma would disrupt  since it would increase the internal inductance at a given edge $q$. \color{black}  
 
 The solenoidal flux $\Psi_{sol}(t)$ can be directly controlled, but this does not give direct control over the plasma-produced flux that lies outside of the plasma,  $\Psi_p^{ex}$.  In the circular-surface model, $\Psi_p^{ex}$ is determined by the total plasma current independent of the profile of that current.    One of the many errors in the Nuclear Fusion paper \cite{Fitzpatrick:2026} is the failure to recognize that the loop voltage on the outermost magnetic surface is not directly controllable since it involves the time derivative of  $\Psi_p^{ex}$, which is given by $dI_p/dt$, in addition to the solenoidal loop voltage $d\Psi_{sol}/dt$. 

A limitation on the advisable loop voltage is electron runaway, which can occur when $V_\ell$ is larger than the Connor-Hastie \cite{Connor-Hastie} value, Equation (\ref{V_ch}).    Exceeding the Connor-Hastie voltage anywhere in the plasma can result in runaway electrons.  The Nuclear Fusion paper \cite{Fitzpatrick:2026} considers the danger minimal citing results from JET.  

The importance of runaways is determined by the strength of the initial seed of runaway electrons and the magnitude of the plasma current before the disruption.  Not only can electrons on the Maxwellian tail form a seed, but power plants have a source of runaway electrons that is not present in JET: Compton scattering of electrons to MeV energies by X-rays emitted from irradiated walls.  With or without a separatrix, the current carried by runaway electrons increases by approximately a factor of ten per megaampere \cite{Boozer:2018} drop in the current until the full plasma current carried by runaways.  The presence of high-Z impurities can cause a substantial reduction in the current change required for a factor of ten increase in the runaway current \cite{Fulop:2017,Tang:2018}.  The exponentiation of the runaway current is due to a change in the poloidal flux outside of the location of the runaways.  This change has a comparable value throughout the plasma.

Not only must the temperature be carefully ramped down to avoid a disruption during shutdown, but also the plasma density must reduced as rapidly as the current to avoid exceeding the Greenwald density limit for disruptions.  That limit is $n_G \equiv I_p/\pi a^2$ where $n_G$ is the line-average electron density through the plasma core in units of $10^{20}$~m$^{-3}$, $I_p$ is the plasma current in megaamperes, and $a$ is the plasma radius in meters.   That limit can be exceeded under special circumstances in tokamaks.  Hurst et al discuss such cases \cite{Greenwald:2024} and provide a recent review of  the Greenwald limit, but there is little reason to believe the Greenwald limit can simply be ignored.  

Power-plant designs generally envision having a plasma density only moderately below the Greenwald limit, so its avoidance requires pumping the excess density out of the chamber on the timescale of the current reduction.  The particle confinement time is longer than the energy confinement time by approximately an order of magnitude, so the reduction in the plasma density would probably require a timescale $\gtrsim 10\tau_E$.  \color{black} The required reduction in the density places special requirements on the pumps.  During normal operation, the separatrix of a tokamak divertor carries the particle exhaust in to very narrow stripes on the wall where the pump ducts are located.  When the separatrix is lost the  particle exhaust strikes the wall as in a tokamak with a limiter, which need not be near the pump ducts. 

%%%%%%%%%%%%%%%%%%%%%%%%%%%%%%%

%%%%%%%%%%%%%%%%%%%%%%%%%%

\section{Faraday's Law and plasma maintenance \label{Sec:plasma maintenance}}

Faraday's Law and the loop voltage not only have important implications about disruptions and plasma shutdown in tokamak power plants but also on plasma maintenance. It is difficult to have fusion pulses in tokamaks longer than approximately a half hour. 

In periods in which a plasma equilibrium is time independent, the magnetic flux enclosed by the axis is
\begin{equation} \Psi_p^{ax}=\Psi_p + \Psi_{sol}(t),
\end{equation} where $\Psi_p$ is time independent.  $\Psi_p$ is the flux enclosed by the magnetic axis that is proportional to the plasma current $I_p$.   The loop voltage on all the magnetic surfaces satisfies $V_\ell(\psi) = d \Psi_{sol}/dt$, a constant.  The profile of the net parallel current $\bar{j}_{||}(\psi_t)$ is given by the constraint of the spatial constancy $V_\ell$.  Since fusion power plants are envisioned to produce their power in steady-state periods, qualitatively, $\eta(\psi_t)\Big(\bar{j}_{||}(\psi_t) -  j_{cd}(\psi_t)-j_{bs}(\psi_t) \Big)$ must be independent of $\psi_t$ during the primary periods of fusion power production. 

The magnitude of the bootstrap current at the magnetic axis is generally negligible.  Without current drive at the axis, the length of steady-state period of tokamak operation is limited by
\begin{equation}
\tau_{max} = \frac{(\Psi_{sol})_{max}-\Psi_p}{2\pi R (\eta \bar{j}_{||})_{ax}}, \label{tau_max}
\end{equation}
where $(\Psi_{sol})_{max}$ is the maximal flux swing of the central solenoid and $(\eta \bar{j}_{||})_{ax}$ is evaluated at the magnetic axis for the planned equilibrium plasma state.  A typical answer for power-plant level tokamaks is that $\tau_{max}$ is approximately a half hour.
 
\color{black}%%%%%%
For longer time-independent periods than $\tau_{max}$, either a stellarator or current drive at the magnetic axis is required.  In 2015 the ARC tokamak power plant was envisioned \cite{ARC:2015} as being steady state, but the present ARC design \cite{ARC2026} assumes 15-minute pulses without current drive. \color{black}  The reason for the change is the intrinsic inefficiency of current drive \cite{CD-eff}.  
\color{black} %%%%%%%%%%

Why is current drive intrinsically inefficient?  Background electrons exert a drag force on the current-carrying electrons, which is minimized when the carriers are mildly relativistic.  The minimal drag force is  $eE_{ch} \approx 0.075 n_{20}$ in International System (SI) Units, where $E_{ch}$ is the Connor-Hastie electric field \cite{Connor-Hastie} and $n_{20}$ is the number density of background electrons in units of $10^{20}/$m$^3$.  The Connor-Hastie loop voltage is
\begin{equation}
V_{ch}=2\pi R E_{ch} \label{V_ch}.
\end{equation}
The power required to drive the full plasma current  \cite{CD-eff} must exceed $I_pV_{ch}$.  

\color{black} %%%%%%%%%%%%%

The 2026 ARC design \cite{ARC2026} has a major radius of $R=4.62$~m, an average electron density of $2.44\times10^{20}\mbox{m}^{-3}$, a net plasma current of $I_p = 12.0$~MA, and a total fusion power of of 1130~MW, which implies an $\alpha$-heating power of 1130/5 = 226~MW.  The theoretical minimum power to drive the full current  $I_pV_{ch}\approx 50$~MW, which is about a quarter of the alpha heating power.  The actual fusion power that would needed to directly drive the full plasma current is several times larger when realistic efficiencies of the whole process are taken into account as well as the density where the current is flowing being somewhat higher than the average density.

\color{black}

%As an example, the ARC V2B design has a major radius of $R = 4.25$~m, a central density of $3.7\times10^{20}\mbox{m}^{-3}$, a net plasma current $I_p = 10.95$~MA, and a total fusion power of 781~MW, which implies an $\alpha$-heating power of 781/5 = 156~MW.  The theoretical minimum power to drive the full current  $I_pV_{ch}\approx 81$~MW, which is over 50\% of the $\alpha$ heating power.  The actual fusion power that would needed to directly drive the full plasma current is several times larger than $I_pV_{ch}$ when realistic efficiencies of the whole process are taken into account as well as the density where the current is flowing being somewhat higher

The inefficiency of current drive implies that a steady-state tokamak must have most of its current produced by the bootstrap effect, but as discussed in Section \ref{Sec: avoidance}, strong bootstrap currents are associated with neoclassical tearing modes, which are subtle to stabilize in a power plant.  Economic fusion is often said to allow only 5\% of the energy production to be used internally in the power plant.

When the total bootstrap current $I_{bs}=\int (j_{bs}/B)d\psi_t$ is significantly greater than the plasma current $I_p$, there is in principle another steady-state, but time dependent, possibility that does not require current drive.  When three conditions are satisfied---$I_{bs}$ is sufficiently large,  the loop voltage at the axis is given by $2\pi R (\eta j_{||})_{ax}$, and $d\Psi_{sol}/dt=0$---the current profile will evolve toward a hollow state---having a local minimum at the axis.  If this profile becomes tearing-mode unstable and forms a central chaotic region, the current profile will flatten preserving magnetic helicity, which is consistent with the constancy of $\int \psi_p d\psi_t$ integrated over the chaotic region, Appendix \ref{Sec:Helicity}.  A major issue is whether cyclic hollowing and flattening can occur without triggering a disruption.

Without strong central current drive supplemented by a large bootstrap current, a tokamak power plant can operate only in short pulses $\lesssim 0.5~$hours, which includes the time required for plasma shutdown and restart.    

In principle, the restart period can be made arbitrarily short by clever choices of the time dependence of $\Psi_{sol}(t)$ and the heating power.   Of course, the periods in which substantial heating power is required must be sufficiently short compared to the periods of fusion burn to have adequate energy to sell.

%%%%%%%%%%%%%%%%%%%%%%%%%%%%

\section{Discussion \label{Sec:Discussion} }

Analytic theory, using only Faraday's Law and mathematics, provides strong constraints on tokamak power plants.  The constraints that follow from Equation (\ref{flux-ev}), the relation between the time derivative poloidal flux and the loop voltage,  are modified by the the breakup of magnetic surfaces but do not disappear.  First, the constraints still apply to all magnetic surfaces that enclose a fixed toroidal flux $\psi_t$ that remain intact.  Second, even when magnetic surfaces are destroyed throughout a volume within the plasma, the flux evolution constraint is replaced by a helicity conservation constraint, which is discussed in the Appendix.

The Faraday-Law constraints raise concerns and should be studied using computational simulations from startup to shutdown of SPARC, ARC, and STEP plasmas.  The constraints could also be studied using existing tokamak experiments to simulate power-plant conditions: input power heating electrons that  has the  spatial and temporal dependence of fusion power, proportional to $(nT_i)^2$,  time constants for changes in low Fourier components of $\vec{B}\cdot\hat{n}$ and the external loop voltage that are orders of magnitude longer than disruption timescales, and limited diagnostics.

Major advances have been made in both stellarator and tokamak power-plant designs by computer optimization.  For stellarators, the focus of optimization is on the external magnetic field.  The external magnetic field defines the  properties of stellarator plasmas more than the properties that can be externally controlled define the properties of any other fusion concept, magnetic or inertial.   For tokamaks, the plasma-profiles are optimized. 

In addition to the extent to which the external magnetic field controls the plasma properties, there are two reasons for this difference in focus in stellarator and tokamak optimization.  First, the number of external magnetic field distributions that are consistent with coils is an order of magnitude larger for stellarators than tokamaks---so large that a full exploration is impossible.  Due to the non-zero toroidal mode number of stellarators, the number of field distributions is approximately $3.6 M_d^2$ times larger than the $M_d$ distributions available in tokamaks \cite{Boozer:RMP}.  Tokamak optimization has focused on three distributions, $M_d=3$---the ones up to triangularity---the equivalent number in a stellarator is thirty-two.  The number of available design options is approximately factorially dependent on the number of field distributions, which reaches more than $10^{35}$ in a stellarator.  Second, a tokamak, especially in a power plant, is in a self-organized state that is largely determined by microturbulent transport processes \cite{Holland:2021}.  The resulting profiles may be consistent or inconsistent with power plants that are disruption free.  

Profile optimization allows the definition of attractive tokamak power plants.  It is concerning when it is not explained how these profiles are to be obtained and controlled in a power plant.  Without profile control, parameters that define not only the current profile but also those that control a fusion burn, such as $n\tau_ET$, could degrade on a shorter timescale than that required for a fusion pulse.  Although not fully understood, the $n\tau_ET$ values in high performance tokamak experiments degrade on a 10~second timescale, but not in stellarators \cite{Long pulse:2024,Long puse:2026}.  Power plants have far stricter limits on both external heating and current drive than existing experiments, which limits the options for control.  

When the objective is purely a demonstration of ignition in a DT plasma, a tokamak with a pulse longer than approximately 10 seconds is adequate, even if followed by a disruption that causes only rapidly repairable damage.  When the objective is the demonstration of the feasibility of tokamaks power plants, the situation is more subtle.

%%%%%%%%%%%%%%%%%%%%%%%%%%%%%%%%%%%%%%%%%%%%%%%%%%%%%%%
\section*{Acknowledgements}

This work received no external support.

 \vspace{0.01in}

\section*{Author Declarations}

The author has no conflicts to disclose. \vspace{0.01in}

%%%%%%%%%%%%%%%%%%%%%%%%%%%%%%%%%%%%%%%%%%%%%%%%%%%%%%%%%%%%%%%%

\section*{Data availability statement}

Data sharing is not applicable to this article as no new data were created or analyzed in this study.

%%%%%%%%%%%%%=============================================

\appendix

%%%%%%%%%%%%%%%%%%%%%%%%%%%%%%%%%===============================

\section{Magnetic Helicity \label{Sec:Helicity} }

When magnetic surfaces are broken and the magnetic field lines become chaotic, the helicity $K$, Equation (\ref{K-def}), not the poloidal flux, becomes the quantity of primary importance.  Chaotic magnetic field lines are defined \cite{Boozer:Chaos} by the infinitesimal separation between neighboring lines having an exponential dependence on the distance along the lines throughout a volume.   

The representation of the magnetic field,
\begin{equation}
2\pi \vec{B} = \vec{\nabla}\psi_t \times \vec{\nabla}\theta + \vec{\nabla}\varphi \times \vec{\nabla}\psi_p(\psi_t,\theta,\varphi)
\end{equation}
 is valid whether the magnetic field is chaotic or not \cite{B-contra}.  Complications due to the guage freedom of the vector potential are eliminated from $dK/dt$ when the integration volume is bounded by perfectly conducting surfaces.   During fast phenomena, such as disruptions, both the magnetic surfaces that remain intact and the chamber walls can be approximated as perfect conductors.

The magnetic helicity $K$ between two perfectly conducting magnetic surfaces $\psi_{in}$ and $\psi_{out}$ is defined by
\begin{eqnarray}
K &\equiv& \int \vec{A}\cdot\vec{B} d^3 x \label{K-def}\\
&=&\int \Big( \psi_t \frac{\partial\psi_p}{\partial\psi_t} -\psi_p\Big)\frac{d\psi_t d\theta d\varphi}{(2\pi)^2} \\
&=& \psi_t\psi_p\Big]_{\psi_{in}}^{\psi_{out}} - 2 \int_{\psi_{in}}^{\psi_{out}} d\psi_t \oint\frac{ d\theta d\varphi}{(2\pi)^2} \psi_p  \hspace{0.2in}\\
\frac{dK}{dt} &=& - 2  \int_{\psi_{in}}^{\psi_{out}} d\psi \oint\frac{ d\theta d\varphi}{(2\pi)^2} \frac{\partial\psi_p}{\partial t}. \label{dK/dt}
\end{eqnarray}

Two situations are of interest:  (1) The chaotic region is bounded by both an inner and an outer perfectly conducting surface.  (2) There is only an outer perfectly conducting surface, which could be the chamber walls as during a major disruption, that bounds the chaotic region.  The second or disruption case is of primary interest.  The current profile in a chaotic region is determined by $j_{||}/B$ being a spatial constant with that constant set by the helicity remaining unchanged.  When the pre-disruption profile of the net parallel current is parabolic $j_{||} \propto (1 -\psi_t/\psi_a)$ with  a conducting wall at the plasma boundary $\psi_a$, then the poloidal flux within the plasma drops to 2/3 of its pre-disruption value, but the plasma current increases by a factor of 4/3 from its strength before the disruption. 

When the magnetic axis is unbroken, the poloidal flux enclosed by the axis is unchanged by a rapid helicity-conserving interaction, nor is the poloidal flux changed through the hole of the torus defined by a magnetic surface that remains intact.

The magnetic helicity in the region between two perfectly conducting surfaces is well preserved even when the plasma is turbulent, which implies the magnetic field lines are chaotic.  Helicity dissipation is given by $2\int \vec{E}\cdot\vec{B}d^3x$ in that region while the magnetic energy is dissipation is given by $\int \vec{E}\cdot\vec{j}d^3x$.  Narrow current channels can dissipate the magnetic energy arbitrarily rapidly, but they produce little helicity dissipation \cite{Taylor:1974,Berger:1984}.

The properties of magnetic helicity $K$ between two perfectly conducting magnetic surfaces $\psi_{in}$ and $\psi_{out}$ are given by Equations (\ref{K-def}) through (\ref{dK/dt}).  The inner and outer magnetic surfaces have skin currents when the chaotic region is an annulus $\psi_{in}<\psi_t<\psi_{out}$.   When $\psi_{in}=0$, there is only one skin current, the one on $\psi_{out}$.  This constraint is trivially satisfied by adding a constant to $\psi_p$ so it equals $\psi_p$ at $\psi_{out}$.  When $\psi_{in}>0$, a surface current flows on the inner surface, which modifies the flux within the annulus $\psi_{in}<\psi_t<\psi_{out}$.

The most important and simplest case is $\psi_{in}=0$ and $\psi_{out} = \psi_a$ the plasma outer boundary.  This will be studied in a cylindrical model as an illustration.  This is a model of a tokamak disruption.   The constancy of $K$ is enforced when $\int \psi_p(\psi) d\psi$ is held constant. 

Assume the pre-disruption current profile is parabolic, $j_{||}\propto (1-r^2/a^2)$, so the enclosed current 
\begin{eqnarray} 
I(r) &=& I_p \frac{4}{b^2}\int_0^r \left(1-\frac{r^2}{a^2}\right) rdr \\
&= & I_p \Big( 2 \frac{r^2}{a^2} -  \frac{r^4}{a^4} \Big) \label{internal L}\\
\psi^{pre}_p(r) &=& 2\pi \mu_0 R \int_0^r \frac{\mu_0 I(r)}{2\pi r} dr \\
&=& \mu_0R I_p \Big(\frac{r^2}{a^2} - \frac{r^4}{4a^4} \Big) \mbox{   or  }\\
\psi^{pre}_p(\psi_t) &=& 2\pi \mu_0 R \Big(\frac{\psi_t}{\psi_a} - \frac{\psi_t^2}{4\psi^2_a} \Big)\\
\int_0^{\psi_b}\psi_p^{pre} d\psi_t &=& \frac{5}{12} \mu_0R I_p\psi_a.
\end{eqnarray}

A single field line comes arbitrarily close to every point in a chaotic region, which implies \cite{j-flat} that on the timescale of a shear Alfv\'en wave, $j_{||}/B$ will be independent $\psi_t$ for $\psi_t<\psi_b$.  The resulting poloidal flux that equals $\psi_p(\psi_a)$ at $\psi_{a}$ is
\begin{eqnarray}
\psi_p^{aft}(\psi) &=& \frac{1}{2}\mu_0R \Big(\frac{ j_{||}}{B}\Big)_c(\psi_t - \psi_{a}) + \psi_p(\psi_a), \hspace{0.3in} \end{eqnarray}
where $\psi_p(\psi_a) = (3/4)2\pi \mu_0 R_{ax}$.
\begin{eqnarray}
\int_0^{\psi_a} \psi^{aft}_p(\psi_t) d\psi_t &=& \int_0^{\psi_a} \Big\{\frac{1}{2}\mu_0R \Big(\frac{ j_{||}}{B}\Big)_c (\psi_t - \psi_{a}) \nonumber\\&& +\frac{3}{4}\mu_0R I_p\Big\}d\psi_t\hspace{0.2in} \\
 &=&-\frac{1}{4}\mu_0R \Big(\frac{ j_{||}}{B}\Big)_c \psi_a^2 \nonumber\\&& +\frac{3}{4}\mu_0R I_p \psi_a\hspace{0.2in}
\end{eqnarray}

Equating the integral $\int \psi_p d\psi_t$ after with that before the disruption
\begin{eqnarray}
-\frac{1}{4}\mu_0R\Big(\frac{ j_{||}}{B}\Big)_c \psi_{a} &=& \Big(\frac{5}{12} -\frac{3}{4}\Big)\mu_0R I_p\\
&=& -\frac{1}{3}\mu_0R I_p  \mbox{   and  }\\
\psi_p^{aft} &=& \left(\frac{1}{4} +\frac{\psi_t}{2\psi_a} \right)\mu_0R I_p. \hspace{0.3in}
\end{eqnarray}

The additive constant on the poloidal flux was chosen so that before the disruption the poloidal flux on axis was zero; afterwards it was $(1/3)\psi_p(\psi_{a})$.  The poloidal flux contained in the plasma changed from $(3/4)\mu_0R I_p$ to $(1/2)\mu_0R I_p$.  The current afterwards $I_{p}^{aft}= (j_{||}/B)_c \psi_a = (4/3) I_p^{pre}$ goes up despite the poloidal flux in the plasma going down.

%%%%%%%%%%%%%%%%%%%%%%%%%

%============================================================================

\end{document}